\documentclass[11pt]{article}

\usepackage[T1]{fontenc}
\usepackage[utf8]{inputenc}
\usepackage{amsmath,amssymb}
\usepackage{graphicx}
\usepackage{xspace}
\usepackage{cite}
\usepackage[a4paper,nohead]{geometry}
\usepackage[bookmarks=true,bookmarksopen=true,colorlinks=true,breaklinks=true,linkcolor=blue,citecolor=blue]{hyperref}
\usepackage[font=small,format=plain,labelfont=bf,up,textfont=it,up]{caption} 
\usepackage[affil-it,max2]{authblk}

\hypersetup{%                                                                
  pdftitle = {Asymptotics of solutions of a hyperbolic formulation of the constraint equations},
  pdfsubject = {},
  pdfauthor = {F. Beyer, L. Escobar, J. Frauendiener},
  pdfkeywords = {}%                                                            
}

\newcommand{\Eqref}[1]{Eq.~\eqref{#1}}
\newcommand{\Eqsref}[1]{Eqs.~\eqref{#1}}
\newcommand{\Sectionref}[1]{Section~\ref{#1}}

\newcommand{\Figref}[1]{Fig.~\ref{#1}}
\newcommand{\Figsref}[1]{Figs.~\ref{#1}}
\newcommand{\s}{\hspace{0.1cm}}

\newcommand{\R}{\ensuremath{\mathbb R}\xspace}

\newcommand{\St}{\ensuremath{\mathbb{S}^2}\xspace}

\newcommand{\mbar}{\overline{m}}

\newcommand{\df} {\mathrm{d}}

\title{Asymptotics of solutions of a hyperbolic formulation of the constraint equations} 

\author[1]{F. Beyer\footnote{Email: fbeyer@maths.otago.ac.nz.} }
\author[2]{L. Escobar\footnote{Email: escobar@math.uni-tuebingen.de.} }
\author[1]{J. Frauendiener \footnote{Email: joergf@maths.otago.ac.nz.} }
\affil[1]{Department of Mathematics and Statistics, University of Otago,  New Zealand.}
\affil[2]{Department of Mathematics, University of T\"ubingen,  Germany.}

\date{\today}

\graphicspath{{plots/}}

\begin{document}

\maketitle

%%%%%%%%%%%%%%%%%%%%%%%%%%%%%%%%%%%%%%%%%%%%%%%%%%%%%%%%%%%%%%%%%%%%%%%%%%%%%%%%%%%%%%%%%%%%%%%%%%%%%%%%%%%%%%%%%%%%%%%%%%%%%%%%%
\begin{abstract}
  In this paper we consider the hyperbolic formulation of the constraints  introduced by R\'acz. Using the numerical framework recently developed by us we construct initial data sets which can be interpreted as nonlinear perturbations of Schwarzschild data in Kerr-Schild coordinates and  investigate their asymptotics. Our results suggest that, unless one finds a way to exploit the freedom to pick the free part of the initial data in some suitable way, generic initial data sets obtained by this method may violate fundamental asymptotic conditions.
\end{abstract}
%%%%%%%%%%%%%%%%%%%%%%%%%%%%%%%%%%%%%%%%%%%%%%%%%%%%%%%%%%%%%%%%%%%%%%%%%%%%%%%%%%%%%%%%%%%%%%%%%%%%%%%%%%%%%%%%%%%%%%%%%%%%%%%%%

\section{Introduction}
The initial value problem plays a hugely important role on the seemingly eternal quest through the largely unmapped terrain of solutions of Einstein's equations. In its most common formulation, see \cite{Friedrich:2000wp,Alcubierre:Book,Ringstrom:2009cj}, one performs a $3+1$-split and thereby decomposes the spacetime and all corresponding geometric tensor fields into spatial and timelike components. As a consequence Einstein's equations naturally split into constraints and evolution equations. 
Thanks to the ground-breaking work by Choquet-Bruhat et al.~\cite{FouresBruhat:1952ji,ChoquetBruhat:1969cl} we know that each solution of the constraints determines a unique solution of the Einstein's equations -- a so-called \emph{maximal globally hyperbolic development} -- in which the initial data set arises as the induced geometry of some spacelike surface. 
In all of what follows, any triple $(\Sigma, h_{ij} , \chi_{ij} )$ of a $3$-dimensional differentiable manifold $\Sigma$, a Riemannian metric $h_{ij}$ and a smooth symmetric tensor field $\chi_{ij}$ on $\Sigma$ is called an \emph{initial data set} if it satisfies the (vacuum) constraint equations 
\begin{equation}\label{StandardConstraints}
{ }^{(3)} R + \chi^2 - \chi_{ij} \chi^{ij} = 0, \quad D^j \chi_{j i} - D_i \chi =0, 
\end{equation}
everywhere on $\Sigma$.
The quantities ${}^{(3)} R $ and $D_i$ are the Ricci scalar and the covariant derivative operator associated with $h_{ij}$, respectively, and $\chi = \chi^{j}_{\ j }$. For this whole paper we agree that operations involving abstract spatial indices $i,j,\ldots$ are performed with $h_{ij}$. 

One easily confirms that the constraints are under-determined in that there are far more degrees of freedom in the two tensor fields $h_{ij}$ and $\chi_{ij}$ than there are constraint equations. 
The constraints therefore leave us with many different ways to separate the free part of the initial data set from the part which we want to determine by solving the constraints. There are thus many quite distinct formulations of the same equations. It turns out that the particular PDE character and hence the  PDE problem naturally associated with it (boundary value problem, initial (boundary) value problem etc.) largely depends on such choices.  The most common formulations of the constraints in the literature are based on the Lichnerowicz-York conformal method, see for instance \cite{Bartnik:2004wn,Baumgarte:2010vs,Isenberg:2014dd}. There are several variations, but all these formulations have in common that they lead to elliptic equations which one then attempts to solve as boundary value problems.

While this is mathematically convenient, and also seemingly natural as a procedure to construct initial data sets, it was realized that the resulting initial data sets can contain ``junk radiation'' and other non-physical features. In numerical relativity one commonly works with initial data sets for which the spatial metric, which is part of the free data in this formulation up to a conformal factor, is conformally flat. However, it was shown \cite{Garat:2000fb} that \emph{rotating} black hole spacetimes do not admit conformally flat slices. This discrepancy is one reason for the observed undesired spurious gravitational wave content  (see, for instance \cite{Alcubierre:Book,Cook:2000dj}). In the literature one can find several attempts to go beyond conformal flatness within the conformal formulation of the constraints, for example, by Matzner, Huq and Shoemaker \cite{Matzner:1998hv} where the free part of the data is determined from a superposition of boosted Kerr–Schild metrics \cite{Cook:2000dj}.

In general, however, it is unclear how to address this obvious discrepancy between  physical meaningfulness and mathematical convenience given that geometric and gauge degrees of freedom are intermixed in a highly nonlinear and largely obscure manner. One possible approach is to study several different formulations of the constraints, construct solutions and then compare the results in order to see if there is one method which stands out both mathematically and physically. One would therefore like to  consider approaches which differ significantly from the standard one above. 

In fact, one sees easily that it is not necessary to formulate the constraints as purely elliptic PDEs. Breaking the restrictive rigidity of elliptic PDEs was for instance a crucial step in the work by Corvino \cite{corvino00:_scalar_einst} who casted the constraints partly into a different form.
Early works where the constraints were formulated as fully non-elliptic \emph{evolution problems} -- i.e., initial value problems as opposed to boundary value problems -- are \cite{Bishop:1998cb,Bartnik:1993er}.
Recently, R\'acz  has brought this idea to a new level in a series of papers \cite{Racz:2014kk,Racz:2014uh,Racz:2015gb}. He proposes two different formulations of the constraints, one where the equations are strongly hyperbolic, and one where they take the form of a parabolic-hyperbolic system (at least under certain restrictions). In the present paper we focus on the former whose details are discussed in \Sectionref{sec:Setup}. The fact that the constraints here are solved  as an \emph{initial value problem} breaks with standard methodologies and interpretations (and maybe even ideologies) which had been developed for the constraints over many decades. Similar to the $3+1$-decomposition of spacetime and associated evolutions with respect to some foliation with spacelike $3$-dimensional surfaces, it is the initial hypersurface $\Sigma$ which is decomposed in a $2+1$-fashion here. The constraints are then solved as an evolution problem with respect to a foliation with $2$-dimensional surfaces. For most of this paper we have in mind that these $2$-dimensional hypersurfaces are spheres, that the \emph{initial} one of these spheres is \emph{finite}  and that  the constraints are then evolved outwards radially towards infinity. This sounds convenient, for example, because numerical codes, in particular \cite{Beyer:2014bu,Beyer:2015bv,Beyer:2016fc,Beyer:2017jw}, which had been developed to solve the evolution equations of Einstein's equations, can be used  without major changes here as well. However, it also has an immediate draw-back: In contrast to the boundary value problem we only have little control over the behavior of the solutions at infinity. In particular, it is, as a matter of principle, impossible to \emph{impose} that the resulting initial data set should be asymptotically flat, asymptotically hyperbolic or whatever asymptotic behavior is of interest. As a consequence of this lack of control, one may therefore end up with initial data sets which are  physically irrelevant. Here we focus on exactly this problem and analyze the asymptotics of non-trivial initial data set families obtained by this method. We restrict to the asymptotically flat case.

There are several notions of \emph{asymptotic flatness} in the literature most of which are motivated by either mathematical or physical considerations (or by both, see  \cite{geroch1972structure, Wald:1984un}). Asymptotically flat initial data sets are supposed to represent a moment of time for isolated systems whose gravitational field decays towards infinity in an appropriate manner. The strongest notion of asymptotic flatness, which we shall consider here, is taken from \cite{Dain:2001cd}. We say that an initial data set $(\Sigma, h_{ij},\chi_{ij})$ is \emph{strongly asymptotically flat} if $\Sigma$ is diffeomorphic to $\R^3$ (possibly minus a ball of finite size) and  if there exist coordinates $\{\tilde{x}^{i}\}$ on $\Sigma$ such that the components of $h_{ij}$ and $\chi_{ij}$ with respect to these coordinates satisfy\footnote{We use the $O$ symbol rather informally in the usual sense $f = O(g) \iff f/g < \infty$ in the relevant limit. Similarly, $f=o(g) \iff f/g \to 0$. In this paper, we avoid the technicalities of norms which must be used to make the definition of the $O$-symbols precise. In fact, in order to make the above notions of asymptotic flatness precise \emph{and} physically meaningful, the $O$-symbol must be defined with respect to a norm which does not only control the decay of the fields themselves, but also guarantees a sufficient decay speed of an appropriate number of derivatives. If this is the case, ``asymptotic flatness'' in the sense above can be shown to imply that the curvature tensor also decays at infinity with some known rate. The interested reader can find the details in the references above.}
\begin{eqnarray}
  h_{ij}    &=& \left(1+\dfrac{2 M}{\varrho}\right) \delta_{ij} + {O}(\varrho^{-2}), \label{eq:AFStrong1}\\
  \chi_{ij} &=& {O}(\varrho^{-2}),\label{eq:AFStrong2}
\end{eqnarray}
in the limit 
\begin{equation}\label{eq:AFLimit}
\varrho := \sum \limits_{i=1}^{3} (\tilde{x}^{i})^2\rightarrow\infty.
\end{equation}
Here $\delta_{ij}$ denotes the Euclidean metric in Cartesian coordinates and $M\ge 0$ is the \emph{ADM mass}. This notion of asymptotic flatness is motivated by the fact that it allows to conformally compactify the initial data set in a way where $\varrho=\infty$ is represented by a single point $i^0$ on the compactified manifold and the fields $h_{ij}$ and $\chi_{ij}$ extend smoothly through $i^0$ after certain conformal rescalings. This is necessary to apply the extraordinarily detailed analysis by Friedrich \cite{friedrich98:_gravit_fields} which yields conditions for which the corresponding solution of the (full) Einstein equations has a smooth conformal compactification. This is a crucial physical property of spacetimes describing isolated systems. In fact, the entire theory of gravitational radiation in general relativity (beyond the linearized level) relies on the existence of such structures.

We refer to such initial data as \emph{strongly} asymptotically flat because there are also weaker notions in the literature. According to \cite{Bartnik:1986dq} the ADM mass of an initial data set is defined if the initial data set is \emph{weakly asymptotically flat}, i.e., if
\begin{eqnarray}
  \label{eq:AFWeak1}
  h_{ij}    &=& \delta_{ij} + {o}(\varrho^{-1/2}), \\
  \chi_{ij} &=& {o}(\varrho^{-3/2}). \label{eq:AFWeak2}
\end{eqnarray}
Spacetime developments determined by such weakly asymptotically flat initial data sets close to the trivial data set were studied in quite some detail in \cite{Bieri:2009tj,Bieri:2009vt}. The assumption of weak asymptotic flatness in these works was a significant improvement over the hypotheses in the original ground-breaking work by Christoudoulou and Klainerman \cite{Christodoulou:1994vm} regarding the nonlinear stability of Minkowski spacetime. In any case, while the extremely difficult and subtle achievements in these works allow to control the decay of fields at infinity just enough to prove stability, only weak conclusions can be drawn regarding the global conformal structure and therefore regarding gravitational radiation. 

In summary it is fair to say that it depends on the particular situation, application and interest whether one wants to consider the strong notion of asymptotic flatness or rather the the weak version above. We are therefore motivated to investigate both here.

The main result of our paper is evidence that, at least in the particular setting we consider here, asymptotically flat initial data sets are unstable (in both the weak and the strong sense) within the family of solutions of the hyperbolic formulation of the constraints. It is therefore likely to be difficult to construct asymptotically flat initial data sets by means of the hyperbolic formulation of the constraints by R{\'a}cz in practice because both the free data and the initial data of the initial value problem of the constraints may need to be fine-tuned in some way. We argue that this conjectured instability  has its origin in the existence of spherically symmetric slices in the Schwarzschild spacetime with  ``the wrong mass'' which are not asymptotically flat. All this is explained in detail in \Sectionref{sec:analysisasymptotics}. In any case, our results are not conclusive enough to claim that R{\'a}cz's formulation of the constraints is ``useless in practice''. After all, we only study a particular setting here, and it may be possible that the freedom in choosing the free part of the data can be exploited in a clever way to solve this issue. Our results do confirm however that, without extra care, the lack of a-priori control of the asymptotics of solutions for this formulations of the constraints can have negative consequences and can in particular  lead to non-physical initial data sets. 

We emphasize that our paper is mostly concerned with the \emph{asymptotics} of solutions of the constraints at spatial infinity. The interesting recent findings in \cite{Winicour:2017wr} therefore have no direct consequences for us.

Our paper is organized as follows. In \Sectionref{sec:Setup} we introduce the $2+1$-decomposition and the hyperbolic formulation of the constraint equations. A particularly interesting setting is obtained when the leaves of the foliations are chosen as $2$-dimensional spheres. This is interesting physically, but also turns out to be convenient technically because one can work with the so-called eth-formalism. All this is discussed in \Sectionref{sec:spheresandeth}. In \Sectionref{sec:exactsols} we consider the Kerr and Schwarzschild solutions in Kerr-Schild coordinates. These yield natural families of exact solutions of the constraints which will be a useful foundation for the analysis in the following sections. \Sectionref{sec:codetest} is devoted to numerical tests. The main part is \Sectionref{sec:analysisasymptotics} where we analyze the asymptotics of a certain class of solutions of the constraints by means of a combination of analytic and numerical methods and thereby derive the main conclusions of our paper.

\section{The vacuum constraint equations as a hyperbolic system}
\label{sec:hyperbolicconstraints}

\subsection{Hyperbolic formulation of the constraints}
\label{sec:Setup}
Let us now discuss the particular hyperbolic formulation of the constraints that we consider in this paper using more or less the original notation introduced in \cite{Racz:2015gb}.  Given any initial data set $(\Sigma,h_{ij},\chi_{ij})$ as described above, one starts from a $2+1$-decomposition of $\Sigma$ (in full analogy with standard $3+1$-decompositions of spacetimes). One picks a smooth function $\rho:\Sigma\rightarrow\R$ such that all the $\rho=const$-level sets are smooth hypersurfaces $S_\rho$ in $\Sigma$ and such that the family of all these hypersurfaces is a foliation of $\Sigma$. There is therefore a smooth \emph{lapse} function $\hat{N}$ on $\Sigma$ so that 
\begin{equation}
  \label{eq:deflapse}
  \hat{n}_{i}=\hat{N} D_{i} \rho
\end{equation}
is a unit normal to the surfaces $S_\rho$. Observe that this lapse $\hat N$ has nothing to do with the lapse of the $3+1$-decomposition of spacetime and, in contrast to the $3+1$-decomposition, $\hat n_i$ is clearly always spacelike here (because the metric $h_{ij}$ is Riemannian). Since $\hat N$ does not vanish as a consequence of the above assumptions, we assume, without loss of generality, that $\hat N>0$ everywhere. The induced metric on $S_\rho$ is 
\begin{equation*}
  \hat{\gamma}_{ij} = h_{ij} - \hat{n}_i \hat{n}_j,
\end{equation*}
and
\[\hat{\gamma}^{i}_{\ j} = \delta^{i}_{\ j} - \hat{n}^{i} \hat{n}_{j}\]
 is therefore the map which projects any tensor field on $\Sigma$ orthogonally to a tensor field which is intrinsic to the surfaces $S_\rho$. Any tensor field, which is annihilated by contractions with $\hat{n}_{i}$ or $\hat{n}^{i}$ on each of its indices, is referred to as \emph{intrinsic} (to the surfaces $S_\rho$) in all of what follows.

Let us now perform a standard decomposition of tensor fields on $\Sigma$ with respect to the normal and tangential directions of $S_\rho$. As a result, all of the relevant geometric quantities will be either scalars or intrinsic tensor fields. We stress that such decompositions are always unique. For $\chi_{ij}$ this gives
\begin{equation}\label{ec:mean_curvature_decomposition}
\chi_{ij} = \text{\boldmath$\kappa$}\, \hat{n}_{i} \hat{n}_{j} + 2\, \hat{n}_{(i} \mathbf{k}_{j)} + \mathbf{K}_{ij} ,
\end{equation}
with
\[\text{\boldmath$\kappa$} =  \hat{n}^{i} \hat{n}^{j} \chi_{ij},\quad \mathbf{k}_j =\hat{\gamma}^{n}_{\ j } \hat{n}^{m} \chi_{nm},\quad \mathbf{K}_{ij} =  \hat{\gamma}^{n}_{\ i}  \hat{\gamma}^{m}_{\ j } \chi_{nm}.\] 
Clearly, the  fields $\mathbf{k}_{j}$ and $\mathbf{K}_{ij}$ are intrinsic and $\mathbf{K}_{ij}$ is symmetric.
The quantity $\mathbf{K}_{ij}$ can be decomposed further as
\begin{equation}\label{eq:trace_decomposition}
	\mathbf{K}_{ij} =  \mathring{ K } _{ij} + \dfrac{1}{2} \hat{\gamma}_{ij} \mathbf{K},
\end{equation}
where
\[\mathbf{K}=\mathbf{K}^l_{\ l}, \quad \mathring{ K } _{ij} \hat\gamma^{ij}=0,\]
and where $\mathring{ K } _{ij}$ is symmetric.
Next we decompose $D_i\hat n_j$:
\begin{equation}
  D_i\hat n_j=\hat K_{ij}+\hat n_i \dot{\hat{n}}_j
\end{equation}
where 
\begin{equation}
  \label{eq:2231}
  \hat K_{ij} =  \hat{\gamma}^{n}_{\ i}  \hat{\gamma}^{m}_{\ j } \hat K_{mn},\quad \dot{\hat{n}}_k=\hat n^i D_i\hat n_j.
\end{equation}
Both $\hat K_{ij}$ and $\dot{\hat{n}}_k$ are purely intrinsic and $\hat K_{ij}$ is symmetric. 
By means of standard calculations, see for example \cite{Racz:2015gb}, one demonstrates that the constraint equations \eqref{StandardConstraints} can be expressed in terms of these quantities as follows
\begin{align}
  \label{eq:constr1}
\mathcal{L}_{\hat{n}} \mathbf{K} - \hat{D}^l \mathbf{k}_l  
&=- 2 \dot{\hat{n}}^i \mathbf{k}_{i} + \left( \text{\boldmath$\kappa$} - \dfrac{1}{2} \mathbf{K} \right) \hat{K} -   \mathring{ K } _{ij}  \hat{K}^{ij} =0,\\
\label{eq:constr2}
  \mathcal{L}_{\hat{n}} \mathbf{k}_{i}  
  - \frac{2}{ \mathbf{K} }  \mathbf{k}^{l} \hat{D}_{i}   \mathbf{k}_{l}
  +\frac{\text{\boldmath$\kappa$}}{ \mathbf{K} } \hat{D}_i \mathbf{K}
&=- \dfrac{1}{ 2\mathbf{K} } \hat{D}_i    \kappa_0   
   -\hat{K} \mathbf{k}_{i} - \left( \text{\boldmath$\kappa$} - \dfrac{1}{2} \mathbf{K} \right) \dot{ \hat{n} }_i +  \dot{ \hat{n} }^l \mathring{ K } _{li} - \hat{D}^l \mathring{ K } _{li},
\end{align}
where $\hat{D}$ is the covariant derivative associated with $\hat\gamma_{ij}$, and,
\begin{equation}
  \label{eq:constr3}
  \hat K=\hat {K^i}_i,\quad \kappa_0 = { }^{(3)} R - \mathring{ K } _{ij} \mathring{ K }^{ij},\quad
  \text{\boldmath$\kappa$} = \dfrac{1}{2 \mathbf{K} }	\left( 2 \mathbf{k}^i \mathbf{k}_i - \dfrac{  \mathbf{K} ^2 }{2}  - \kappa_0  \right).
\end{equation}

In order to interpret \Eqsref{eq:constr1} -- \eqref{eq:constr2} with \Eqref{eq:constr3} as a system of evolution equations now, we pick, in addition to the above, a smooth vector field $\rho^i$ normalized as $\rho^i  D_i  \rho = 1$. Effectively, solving this system as an evolution problem means that we shall integrate along this vector field; see below. We find
\begin{equation}
  \label{eq:defrho}
  \rho^i=\hat N \hat n^i+\hat N^i,
\end{equation}
where $\hat N^i$ is an intrinsic \emph{shift} vector field, i.e., $\hat{N}^{i} = \hat{\gamma}^{i}_{\ j} \rho^{j}$; recall the definition of the lapse $\hat N$ in \Eqref{eq:deflapse}. Let us now introduce the following terminology:
\begin{description}
\item[Free data: ] The fields 
\begin{equation}
  \label{eq:freedata}
  \hat{\gamma}_{ij},\quad \hat N,\quad \hat N^i,\quad \mathring{ K } _{ij}\quad\text{and}\quad \hat K_{ij}, 
\end{equation}
introduced above
are considered as \emph{freely specifiable data} on $\Sigma$, i.e., \emph{free data}, for the system \Eqsref{eq:constr1} -- \eqref{eq:constr3}. Observe that the full $3$-dimensional metric $h_{ij}$ is determined by these free data alone. In particular, one can therefore calculate ${ }^{(3)} R$ in \Eqref{eq:constr3} and $\dot{\hat{n}}_k$ according to \Eqref{eq:2231} \emph{before} the system is solved. Given $\hat N$ and $\hat N^i$, the Lie derivatives with respect to $\hat n^i$ in \Eqsref{eq:constr1} and \eqref{eq:constr2} can be written as derivatives along $\rho^i$ (``evolution derivatives'') and derivatives tangential to $S_\rho$. This justifies the intuition of considering our system as an evolution problem where one integrates along the field $\rho^i$.
\item [Unknowns: ] The fields $\mathbf{K}$ and $\mathbf{k}_{i}$ are considered as the \emph{unknowns} of the system \Eqsref{eq:constr1} -- \eqref{eq:constr3}. Once a solution has been found, one can calculate $\text{\boldmath$\kappa$}$ from \Eqref{eq:constr3}, and therefore the complete initial data set $h_{ij}$ and $\chi_{ij}$.
\item[Hyperbolicity condition: ] The inequality
\begin{equation}\label{eq:hyperbolicitycondition}
  \text{\boldmath$\kappa$} \ \mathbf{K} <  0 
\end{equation}  
is referred to as the \emph{hyperbolicity condition}. 
\end{description}
It is shown in \cite{Racz:2015gb} that for any smooth free data \Eqref{eq:freedata} on $\Sigma$, the system \Eqsref{eq:constr1} -- \eqref{eq:constr2} 
with \Eqref{eq:constr3}  is a quasilinear strongly (in fact, strictly) hyperbolic system with respect to the foliation of surfaces $S_\rho$ for 
the unknowns $\mathbf{K}$ and $\mathbf{k}_{i}$ provided that \eqref{eq:hyperbolicitycondition} holds. Therefore, it is natural to refer to such
 an inequality as the \emph{hyperbolicity condition} as above. Observe that the principal part of the system can be written as
\begin{equation*}
  \begin{pmatrix}
    \hat n^j & -\hat\gamma^{jl}\\
    \frac{\text{\boldmath$\kappa$}}{\mathbf{K}}{\delta_i}^j & \hat n^j {\delta_i}^l -\frac{2}{\mathbf{K}}\hat\gamma^{ml}{\delta_i}^j \mathbf{k}_{m} 
  \end{pmatrix} D_j
  \begin{pmatrix}
    \mathbf{K}\\
    \mathbf{k}_{l} 
  \end{pmatrix}.
\end{equation*}
\Eqsref{eq:constr1} -- \eqref{eq:constr2} can therefore always be symmetrized \cite{Racz:2015gb} by multiplying it from the left with the matrix
\begin{equation*}
  \begin{pmatrix}
    -\frac{\text{\boldmath$\kappa$}}{\mathbf{K}} & 0\\
    0 & \gamma^{ik}
  \end{pmatrix}.
\end{equation*}
Since the matrix
\begin{equation*}
  \begin{pmatrix}
    -\frac{\text{\boldmath$\kappa$}}{\mathbf{K}} & 0\\
    0 & \gamma^{ik}
  \end{pmatrix}\cdot \begin{pmatrix}
    \hat n^j & -\hat\gamma^{jl}\\
    \frac{\text{\boldmath$\kappa$}}{\mathbf{K}}{\delta_i}^j & \hat n^j {\delta_i}^l -\frac{2}{\mathbf{K}}\hat\gamma^{ml}{\delta_i}^j \mathbf{k}_{m} 
  \end{pmatrix} \hat n_j
\end{equation*}
is always positive definite provided \Eqref{eq:hyperbolicitycondition} holds, \Eqsref{eq:constr1} -- \eqref{eq:constr2} with \Eqref{eq:constr3} can indeed be transformed into a quasilinear symmetric hyperbolic system with respect to the foliation of surfaces $S_\rho$. In particular, this implies that given any smooth {free data} as in \Eqref{eq:freedata} on $\Sigma$, the \emph{initial value problem} of \Eqsref{eq:constr1} -- \eqref{eq:constr2} with \Eqref{eq:constr3} is well-posed for the class of all smooth \emph{initial data} $\left.\mathbf{K}\right|_{\rho=\rho_0}$ and $\left.\mathbf{k}_{i}\right|_{\rho=\rho_0}$ prescribed on any smooth \emph{initial surface} $S_{\rho_0}$ for which 
\[\left.\left(\text{\boldmath$\kappa$} \ \mathbf{K}\right)\right|_{\rho=\rho_0} <  0.\]
This means that for any such data (free data and initial data) there exists a unique smooth solution $(\mathbf{K}, \mathbf{k}_{i})$ of \Eqsref{eq:constr1} -- \eqref{eq:constr2} with \Eqref{eq:constr3}, at least on some (possibly small) interval $[\rho_0,\rho_1)$.

Suppose for the moment that $\Sigma$ can be identified with $\R^3$ (possibly outside some compact subset) and that the limit $\rho\rightarrow\infty$ describes the geometric asymptotics at spacelike infinity. The well-posedness of the initial value problem of our system, which we have just established under the conditions above, is certainly crucial. However, there are two further questions: (i) Do any of the solutions extend to $\rho\rightarrow\infty$? After all, the hyperbolicity condition \eqref{eq:hyperbolicitycondition} involves the unknowns and could therefore break down at any point of the evolution. (ii) What is the asymptotic geometric behavior of the resulting initial data set? The answers to these questions are completely unclear and will (at least partly) be addressed in this article. 

\subsection{Foliation by spheres and the eth-formalism}
\label{sec:spheresandeth}
For the rest of this paper we shall restrict to the case that $\Sigma$ can be identified with $\R^3$ (possibly minus some compact subset) and that the surfaces $S_{\rho}$ introduced above are diffeomorphic to $2$-spheres for all $\rho$ larger than some $\rho_0>0$. The initial value problem of \Eqsref{eq:constr1} -- \eqref{eq:constr2} with \Eqref{eq:constr3} can then be understood as an evolution problem with ``time'' $\rho$ and where the ``spatial topology'' is $\mathbb S^2$. Some of the technical complications in solving this evolution problem both analytically and numerically arise from the fact that $\mathbb S^2$ cannot be covered by a single regular coordinate chart. The description of tensor fields on $\mathbb S^2$ in terms of any such single coordinate chart breaks down at some ``spatial'' point. This problem is often called  the \textit{pole problem} because in standard polar coordinates for \St these issues appear at the poles. 

This problem is resolved in the so-called \emph{$\eth$-formalism} (or eth-formalism) where all formally singular terms associated with the coordinate singularity at the poles are absorbed into geometrically motivated regular differential operators.
Without going into the details (see~\cite{Penrose:1984tf,Beyer:2014bu,Beyer:2015bv,Beyer:2016fc}), here is a brief summary. One first introduces coordinates $(\rho,\theta,\varphi)$ on $\Sigma$ for $\rho>\rho_0$ which are adapted to the foliation, i.e.,  the field $\rho^i$ in \Eqref{eq:defrho} is identified with the coordinate field $\partial_\rho^i$ and $(\theta,\varphi)$ are standard polar coordinates on each $S_\rho$ (diffeomorphic to $\mathbb S^2$). Given these, one introduces the frame $(m^i,\mbar^i)$ (defined almost everywhere on $S_\rho$ for any $\rho$) where
\begin{equation}\label{eq:referenceframe}
m^i:=\frac 1{\sqrt 2}\left(\partial^i_{\theta}-\frac{\text{i}} {\sin\theta}\partial^i_\varphi\right);
\end{equation}
complex conjugation is denoted by a bar and $\text{i}$ is the imaginary unit. The corresponding dual frame is $(\omega_k,\overline\omega_k)$ where
\begin{equation*}
\omega_k := \frac 1{\sqrt 2} \left( \df \theta_k + \text{i} \sin\theta \, \df  \varphi_k \right).
\end{equation*}
Any tensor field which is intrinsic to $S_\rho$ can be expressed in terms of this frame at each value of $\rho>\rho_0$. Our convention is that the index $1$ refers to the projection onto the $m^i$ or $\omega_i$ direction, while the index $2$ is associated analogously with  $\mbar^i$ or $\overline\omega_i$. For example, we therefore have (at any fixed value of $\rho$)
\begin{equation*}
\hat\gamma_{ij} =   2 \hat\gamma_{12}\, \omega_{(i} \overline\omega_{j)}  + \hat\gamma_{11}\,  \omega_i \omega_j + \hat\gamma_{22}\, \overline\omega_i \overline\omega_j.
\end{equation*}
Observe that $\hat\gamma_{12}$ must therefore be a real function while $\hat\gamma_{11}=\overline{\hat\gamma_{22}}$ are complex functions on $\mathbb{S}^2$. Note, that the metric of the unit-sphere is obtained with $\hat\gamma_{11}=0$ and $\hat\gamma_{12}=1$. Note further, that the metric remains unchanged under \emph{frame rotations} $m^i \mapsto \text{e}^{\text{i} \alpha}\,m^i$ for arbitrary functions $\alpha: \Sigma \to \R$.

Given this frame, the \emph{spin-weight formalism} (see the references above) assigns to any component of a smooth intrinsic tensor field with respect to this frame a well-defined integer \emph{spin-weight}. The spin-weight describes the transformation behavior of the components of tensor fields under pointwise frame rotations of the frame $(m^i,\mbar^i)$ (and correspondingly of $(\omega_i,\overline\omega_i)$) as defined above. By convention, $m^i$ has the spin-weight $1$, $\mbar^i$ the spin-weight $-1$, $\omega_i$ the spin-weight $-1$ and $\overline \omega_i$ the spin-weight $1$. Correspondingly,  the function $\hat\gamma_{12}=\hat\gamma_{ij}m^i\mbar^j$ has the spin-weight $0$, $\hat\gamma_{11}=\hat\gamma_{ij}m^i m^j$ the spin-weight $2$ and $\hat\gamma_{22}=\hat\gamma_{ij}\mbar^i\mbar^j$ the spin-weight $-2$ etc. Any sufficiently regular function $f$ on $\mathbb{S}^2$ (and therefore on $S_\rho$ for any fixed $\rho>\rho_0$) with spin-weight $s$ can be expanded in terms of \emph{spin-weighted spherical harmonics} (SWSH) $_{s}Y_{lm}$ as follows
\begin{equation}\label{eq:functionS2}
f (\theta,\varphi) =  \sum\limits_{l=|s|}^{\infty}  \sum\limits_{m=-l}^{l} {}_{s}a_{lm} \,_{s}Y_{lm} (\theta,\varphi) ,
\end{equation}
for complex coefficients $_{s}a_{lm}$. Thus, a spin-weighted quantity can be represented in terms of local coordinates $(\theta,\varphi)$ (on the left) of in terms of its expansion coefficients ${}_{s}a_{lm}$ (on the right).

Due to the orthonormality relation
\begin{equation}\label{integral_properties_spherical_harmonics}
  \langle \  {}_{s} Y_{l_1 m_1 } \ | \ _{s}{Y}_{l_2 m_2} \ \rangle := \int \limits_{\mathbb{S}^2} \s  {}_{s} Y_{l_1 m_1 }(\theta,\varphi) \: _{s}\overline{Y}_{l_2 m_2}(\theta,\varphi) \s \df\Omega = \delta_{l_1 l_2} \delta_{m_1 m_2},
\end{equation}
the coefficients ${}_{s}a_{lm}$ in \Eqref{eq:functionS2} can be calculated as 
\begin{equation}
  {}_{s}a_{lm}=\langle \  f \ | \ {}_{s}{Y}_{l_2 m_2}\ \rangle.
\end{equation}

If $f$ is a function of spin-weight $s$ on $\mathbb S^2$, the \textit{eth-operators} $\eth$ and $\bar{\eth}$ are  defined as
\begin{equation}\label{eq:def_eths}
\eth f       := \partial_\theta f - \dfrac{\text{i}}{ \sin \theta} \partial_\varphi f - s  \cot \theta\,f, \quad 
\bar{\eth} f := \partial_\theta f + \dfrac{\text{i}}{ \sin \theta} \partial_\varphi f + s \cot \theta\, f.
\end{equation}
Using  \Eqsref{eq:def_eths} and \eqref{eq:referenceframe}, we can therefore express directional derivatives along the frame vectors $(m^i,\mbar^i)$ in terms of the eth-operators as
\begin{equation}\label{eq:ethm}
m( f )=m^i D_i f = \dfrac{1}{\sqrt 2} \left(  \eth f  + f  s \cot\theta  \right), \quad \mbar( f )=\mbar^i D_i f = \dfrac{1}{\sqrt 2} \left( \bar{\eth} f  -  f s \cot\theta   \right) .
\end{equation}
The operator $\eth$ raises the spin-weight by one while $\bar{\eth}$ lowers it by one. Of particular importance are the identities
\begin{equation}\label{eq:eths}
\begin{split}
\eth  \hspace{0.1cm}_{s}Y_{lm} (\theta,\varphi)  &= - \sqrt{ (l-s)(l+s+1) } \hspace{0.1cm}_{s+1}Y_{lm} (\theta,\varphi) , \\
\bar{\eth}   \hspace{0.1cm}_{s}Y_{lm} (\theta,\varphi)   &= \sqrt{ (l+s)(l-s+1) } \hspace{0.1cm}_{s-1}Y_{lm} (\theta,\varphi) , 
\end{split}
\end{equation}
which can also be used to define the spin-weighted spherical harmonics $_{s}Y_{lm}$ for any integer $s$ from the standard spherical harmonics $Y_{lm}={}_{0}Y_{lm}$. It is worthwhile to point out that the power of the $\eth$-formalism arises from the fact that the SWSH and the $\eth$ and $\bar{\eth}$ operator are \emph{globally defined} on $\mathbb{S}^2$. We have introduced them above in terms of their local representations in the usual polar coordinates with respect to the frame \eqref{eq:referenceframe}. This globality and, in particular the raising and lowering properties \eqref{eq:eths}, allow us to compute derivatives and integrals of spin-weighted quantities very easily using fast transformations between their representations with respect to the SWSH and the local coordinates.

In order to apply this framework to \Eqsref{eq:constr1} -- \eqref{eq:constr3} one proceeds as follows. First all intrinsic fields in these equations are expressed in terms of the frame above. Each quantity in the equations is then described in terms of functions on $(\rho_0,\infty)\times\mathbb S^2$ (i.e., they depend on $\rho$, $\theta$ and $\varphi$) each of which has a well-defined spin-weight at each fixed $\rho$ which is constant in $\rho$. Next, all directional derivatives along $m^i$ and $\mbar^i$ are expressed in terms of $\eth$ and $\bar{\eth}$ according to \Eqref{eq:ethm}. After simplifying the resulting equations algebraically, one obtains a system of equations, where all terms in each equation have the same spin-weight and all terms are  explicitly regular (i.e., the pole problem disappears). Because the full set of equations which one obtains are quite lengthy, we will not write them down now; see \cite{Racz:2016wb}. For brevity we only write them down in the special case below in \Eqref{sys:simplifiedsystemofequations}.

\newcommand{\BbbK}{\mathbf{K}}
\newcommand{\Bbbkm}{\mathbf{k}_1}
\newcommand{\Bbbkmbar}{\mathbf{k}_2}

\subsection{Some explicit solutions of the constraints}
\label{sec:exactsols}

This subsection is devoted to some explicit solutions of the constraints \Eqsref{eq:constr1} -- \eqref{eq:constr3}.  On the one hand, we will use these to test our numerical code in \Sectionref{sec:codetest}. On the other hand, however, they will also be the foundation for our analysis of non-trivial nonlinearly perturbed initial data in \Sectionref{sec:analysisasymptotics}.

\paragraph{Kerr-Schild initial data.} We start with the Kerr spacetime \cite{Hawking:1973tb}. As noted before, \Eqref{eq:hyperbolicitycondition} is incompatible with time-symmetric slices. The initial data sets obtained by $t=const$-slices of the Kerr metric in Boyer-Lindquist coordinates are therefore not compatible with our formulation of the constraints. This problem can be reconciled by  considering instead $t=const$-surfaces of the Kerr metric in Kerr-Schild coordinates.  As these slices are not time-symmetric, but asymptotically flat, they shall turn out to be useful for our studies here.

In Kerr-Schild coordinates $(t,x,y,z)$, the Kerr metric with mass $M$ and angular momentum parameter $a$ takes the form \cite{Misner:1973vb,Racz:2015bu}
\begin{equation*}
  g_{\mu \nu } = \eta_{\mu \nu } + 2 H l_{\mu} l_{\nu} , 
\end{equation*}
where 
\begin{equation*}
H =  \dfrac{r M}{r^2 + a^2 z^2 / 2},\quad
l_{\mu} = \df t_\mu+\dfrac{r x + a y} { r^2 + a^2}\df x_\mu +  \dfrac{r y - a x }{ r^2 + a^2}\df y_\mu+ \dfrac{ z }{ r }\df z_\mu,
\end{equation*}
the field $\eta_{\mu \nu}$ is the Minkowski metric in standard Cartesian coordinates $(t,x,y,z)$,
and $r$ is the unique positive real solution of the equation
\begin{equation*}
  \dfrac{ x^2 + y^2 }{ r^2 + a^2 } + \dfrac{z^2}{r^2} = 1 .  
\end{equation*}
The limit $r\rightarrow\infty$ corresponds to spacelike infinity.
Observe that $l_\mu$
is null both with respect to $g_{\mu \nu}$ and $\eta_{\mu \nu}$. 

The induced metric $h_{ij}$ and second fundamental form $\chi_{ij}$ on any $t=const$-surface $\Sigma$ in this spacetime is an initial data set. It can therefore  be expressed as in Sections~\ref{sec:Setup}~and~\ref{sec:spheresandeth}. The $\rho=r=const$-surfaces on $\Sigma$ are diffeomorphic to spheres and provide a foliation of $\Sigma$ for any $r>0$. Therefore, in all of what follows we will write $r$ instead of $\rho$.

The components of the intrinsic metric on  $\St$ are

\begin{align*}
	\hat\gamma_{11} &= -\frac{a^2 \sin ^2\theta  \left(a^2 \cos 2 \theta +a^2+2 r (2 M+r)\right)}{4 \left(a^2 \cos ^2 \theta +r^2\right)}, \\
	\hat\gamma_{12}  &=  \frac{-a^2 \sin ^2\theta \left(a^2+r (r-2 M)\right)+\left(a^2 \cos ^2 \theta +r^2\right)^2+\left(a^2+r^2\right)^2}{2 \left(a^2 \cos ^2\theta +r^2\right)},
\end{align*}
and $\hat\gamma_{22}= \overline{\hat\gamma_{11}}$. The lapse function takes the form
\begin{equation*}
\hat{N}= \frac{\sqrt{a^2 \cos 2 \theta +a^2+2 r^2} \sqrt{a^2 \cos 2 \theta +a^2+2 r (2 M+r)}}{\sqrt{2} \sqrt{a^4+a^2 \cos 2 \theta  \left(a^2+r (r-2 M)\right)+a^2 r (2 M+3 r)+2 r^4}},
\end{equation*}
and the components of the shift vector are
\begin{equation*}
\hat N^1  =	-\frac{	\text{i} a \sin \theta  \left(a^2 \cos 2 \theta +a^2+2 r (2 M+r)\right)}{\sqrt{2} \left(a^4+a^2 \cos 2 \theta \left(a^2+r (r-2 M)\right)+a^2 r (2 M+3 r)+2 r^4\right)},
\end{equation*}
and $\hat N^2=\overline{\hat N^1}$.
The components of $\mathbf{K}_{ij}$ are
\begin{align*}
\mathbf{K}_{11} &= \frac{a^2 M C r \sin ^2\theta \left(-\text{i} a^3 \cos 3 \theta +a^2 M \cos 2 \theta +a^2 M-\text{i} a \left(3 a^2+4 r^2\right) \cos \theta-2 M r^2\right)}{2 \left(a^2 \cos ^2\theta +r^2\right)^2 \left(a^2 \cos ^2 \theta +r (2 M+r)\right)} ,\\
\mathbf{K}_{12} &= - M C r  \Big(-a^4 \cos 4 \theta  (M-2 r)+a^4 M+6 a^4 r-4 a^2 M r^2 +4 a^2 r \cos 2 \theta  \left(2 a^2 \right. \\
 &  \left. +r (M+4 r)\right)  +16 a^2 r^3+16 r^5 \Big)  \bigg/  8 \left(a^2 \cos ^2 \theta +r^2\right)^2 \left(a^2 \cos ^2\theta+r (2 M+r)\right),\\
\end{align*}
with
\begin{equation*}
C= \sqrt{\frac{a^2 \cos 2 \theta +a^2+2 r (2 M+r)}{a^2 \cos 2 \theta +a^2+2 r^2}},
\end{equation*}
and $\mathbf{K}_{11}=\overline{\mathbf{K}_{22}}$.
Via \Eqref{eq:trace_decomposition}, this allows us to calculate the components 
of
$ \mathring{ K }_{ij}$ and of $\BbbK$.
The latter function and the components of $\mathbf{k}_{i}$ are given by
\begin{align}
\label{eq:BbbKKerr}
\BbbK =\,& \frac{4 M r \left(a^2 \cos 2 \theta (M-r)-a^2 (M+3 r)-4 r^3\right)}{\left(a^2 \cos 2 \theta +a^2+2 r^2\right) \left(a^4+a^2 \cos 2 \theta  \left(a^2+r (r-2 M)\right)+a^2 r (2 M+3 r)+2 r^4\right) C}	,\\
\Bbbkm=\,& - \text{i} a M C \sin \theta \Big(3 a^6+a^4 r (8 M-3 r)-24 a^2 r^3 (M+r)-2 \text{i} a r \left(5 a^4+12 a^2 r^2  \right. \nonumber\\
& \left.  +8 r^4\right) \cos \theta  \left. \left.  + a^2 \left(-\text{i} a^3 r \cos 5 \theta -\text{i} a r \left(5 a^2+8 r^2\right) \cos 3 \theta +a^2 (a-r) (a+r) \cos 4 \theta  \right. \right.\right.  \nonumber\\
&  +4 \cos 2 \theta \left.  \left(a^4+a^2 r (2 M-r)-2 r^3 (M+2 r)\right)\right)-24 r^5 (2 M+r) \Big)  \bigg/  \nonumber\\
& \left(D \left(a^2 \cos 2 \theta+a^2+2 r^2\right)^2 \left(a^2 \cos 2 \theta +a^2+2 r (2 M+r)\right)^2\right)\notag,
\end{align}
where
\begin{equation*}
D = \sqrt{\frac{a^2+r^2}{a^2 \cos 2 \theta+a^2+2 r^2}-\frac{2 M r}{a^2 \cos 2 \theta +a^2+2 r (2 M+r)}} 
\end{equation*}
and $\Bbbkmbar=\overline{\Bbbkm}$.

\paragraph{The Schwarzschild case and nonlinear perturbations.} 
The special case of Kerr-Schild Schwarzschild initial data is obtained from the above by setting $a=0$ for any $M\ge 0$. In this case we have
\begin{equation}\label{schwarzschildvaluesinitiadata}
\hat{\gamma}_{ij} = 2 r^2 \omega_{(i} \overline{\omega}_{j)},\quad
\hat{N}  = \sqrt{ 1 + 2 M /r }, \quad
\hat{N}^i= 0 , \quad \mathring{ K }_{ij} = 0.
\end{equation}
Observe that the $S_{r}$ level sets are therefore round spheres with radius $r$. We calculate from \Eqref{eq:constr3} that
\begin{equation}\label{def:KoInSchwarchild}
  \kappa_0 = {}^{(3)} R = \dfrac{8 M^2}{ r^2 (2M+r)^2 }. 
\end{equation}
Moreover, we have
\begin{equation}
\label{eq:SSdata}
\BbbK=-\dfrac{4 M}{r^{3/2} \sqrt{2 M+r}},\quad \Bbbkm=\Bbbkmbar=0.
\end{equation}

In order to define \emph{nonlinear perturbations} of these exact Kerr-Schild Schwarzschild data now we proceed as in \cite{Racz:2015bu}: We keep the exact Schwarzschild quantities \Eqref{schwarzschildvaluesinitiadata} as the free data, but then consider $\BbbK$ and $\Bbbkm$ as the general unknowns of \Eqsref{eq:constr1} -- \eqref{eq:constr3}, which thereby take the form
\begin{equation}\label{sys:simplifiedsystemofequations}
\begin{split}
\partial_r \BbbK &= -\frac{3 \BbbK }{2 r} -\frac{\kappa_0}{r \BbbK } +\frac{4 |\Bbbkm|^2  }{r^3 \BbbK } + \sqrt{\frac{M}{r}+\frac{1}{2}}  \frac{ \ \left( \eth\Bbbkmbar  + \bar{\eth }\Bbbkm  \right) }{r^2}, \\
\partial_r \Bbbkm      &=   -\frac{2 \Bbbkm }{r}  \\
&\quad\, + \sqrt{\frac{M}{r}+\frac{1}{2}} \, \frac{ \eth \BbbK \left(r^2 \left(2 \kappa_0 +\BbbK^2\right)-8 |\Bbbkm|^2 \right)+8 \BbbK \left(\Bbbkm  \eth \Bbbkmbar +\Bbbkmbar \eth \Bbbkm \right)-2 r^2 \BbbK \eth \kappa_0}{4 r^2 \BbbK^2},
\end{split}
\end{equation}
with $\Bbbkmbar=\overline{\Bbbkm}$ and \Eqref{def:KoInSchwarchild}; in applying the eth-operators in these equations, we must make use of the spin-weights $1$ for $\Bbbkm$ and $-1$ for $\Bbbkmbar$ consistently.
 One can show easily \cite{Racz:2015bu} that the hyperbolicity condition is satisfied for \Eqref{sys:simplifiedsystemofequations} for all initial data sufficiently close to data determined by \Eqref{eq:SSdata} on any initial sphere $r=r_0>0$. The initial value problem is therefore well-posed. \emph{The solutions of \Eqref{sys:simplifiedsystemofequations} given by different initial data for $\BbbK$ and $\Bbbkm$  are interpreted as  different nonlinear perturbations of the Kerr-Schild Schwarzschild initial data set}.

One can easily find an explicit family of such nonlinear perturbations by assuming that $\Bbbkm=\Bbbkmbar=0$ 
and that $\BbbK$ only depends on $r$.
In this case, \Eqref{sys:simplifiedsystemofequations} reduces to the single ordinary differential equation 
\begin{equation}
\partial_r \BbbK = -\frac{3 \BbbK }{2 r} -\frac{\kappa_0}{r \BbbK } .\label{eq:appriximation3} 
\end{equation}
Using \Eqref{def:KoInSchwarchild} we obtain the following family of solutions
\begin{equation}\label{ec:schwarchildcase}
  \Bbbkm=\Bbbkmbar=0,\quad \BbbK =  \pm \dfrac{1}{r^{3/2}} \sqrt{ C + \frac{16M^2}{r+2 M} },
\end{equation}
where $C$ is an integration constant. We impose the restriction $C \ge 0$ so that $\BbbK $ is real for all large values of $r$. The case of exact Kerr-Schild Schwarzschild data corresponds to the case $C=0$, and, by choice of the time orientation, to the overall negative sign. In \Sectionref{sec:sphericallysymmdata} we discuss the significantly different properties of the initial data sets obtained for $C=0$ and $C>0$.

\section{Numerical implementation and tests}
\label{sec:codetest}

Since the general hyperbolic constraint equations can be formulated completely in the eth-formalism under the assumptions in \Sectionref{sec:hyperbolicconstraints} (this is not restricted to the special case in \Eqref{sys:simplifiedsystemofequations}), the pseudo-spectral numerical framework developed in \cite{Beyer:2014bu,Beyer:2015bv,Beyer:2016fc,Beyer:2017jw} applies naturally. The  idea is that the unknowns, all of which have a well-defined spin-weight in this formalism, are expanded in terms of spin-weighted spherical harmonics as in \Eqref{eq:functionS2} at any value $r=\rho$. The complex coefficients in the expansions hence only depend on $r$. The ``spatial derivatives'' in the equations can be calculated explicitly using \Eqsref{eq:eths}.
In this paper we restrict to the \emph{axially symmetric case} where we assume that all the unknowns are constant along $\varphi$. In this case, we can use the highly efficient and accurate code developed in \cite{Beyer:2016fc}. We refer the reader to  these references  for details.

The main purpose of this section is to discuss numerical tests using the exact Kerr-Schild Kerr data solution of the constraints for $M=1$ and $a=0.5$ considered in \Sectionref{sec:exactsols}. 
For comparing the numerical and the exact solution, we define 
\begin{equation*}\label{eq:error_definition}
E(r):=  \sqrt{  \| \hspace{0.1cm}   \BbbK^{(num)}  - \BbbK \hspace{0.1cm}  \|^2_{L^2(\St)}  + \| \hspace{0.1cm}  \Bbbkm^{(num)} - \Bbbkm \hspace{0.1cm}  \|^2_{L^2(\St)}  }
\end{equation*}
as our \emph{error measure}.
The quantities $\BbbK^{(num)}$ and $\Bbbkm^{(num)}$ represent the numerical solutions while $\BbbK$ and $\Bbbkm$ represent the exact solution given in \Sectionref{sec:exactsols}. The norm\footnote{For simplicity we are not using the correct $\mathbb S^2$-volume element to approximate this ``norm''; we find that the inclusion of the correct volume element here does not change the results significantly. } $ \| \cdot \|_{L^2(\St)}$  is numerically  approximated  by
\begin{equation*}
\| \:  f  \:  \|_{L^2(\St)} \approx \sqrt{ \dfrac{ 2 \pi^2 }{ N } \:  \sum \limits^{N}_{n=0} \: |f_n|^2 }
\end{equation*}
for any function $f$ sampled on the grid of $N$ points in the $\theta$ direction; recall that due to axial symmetry we do not need to sample in the $\varphi$ direction. 

\begin{figure}[t]
  \begin{minipage}{0.49\linewidth}
    \centering
     \includegraphics[width=\linewidth]{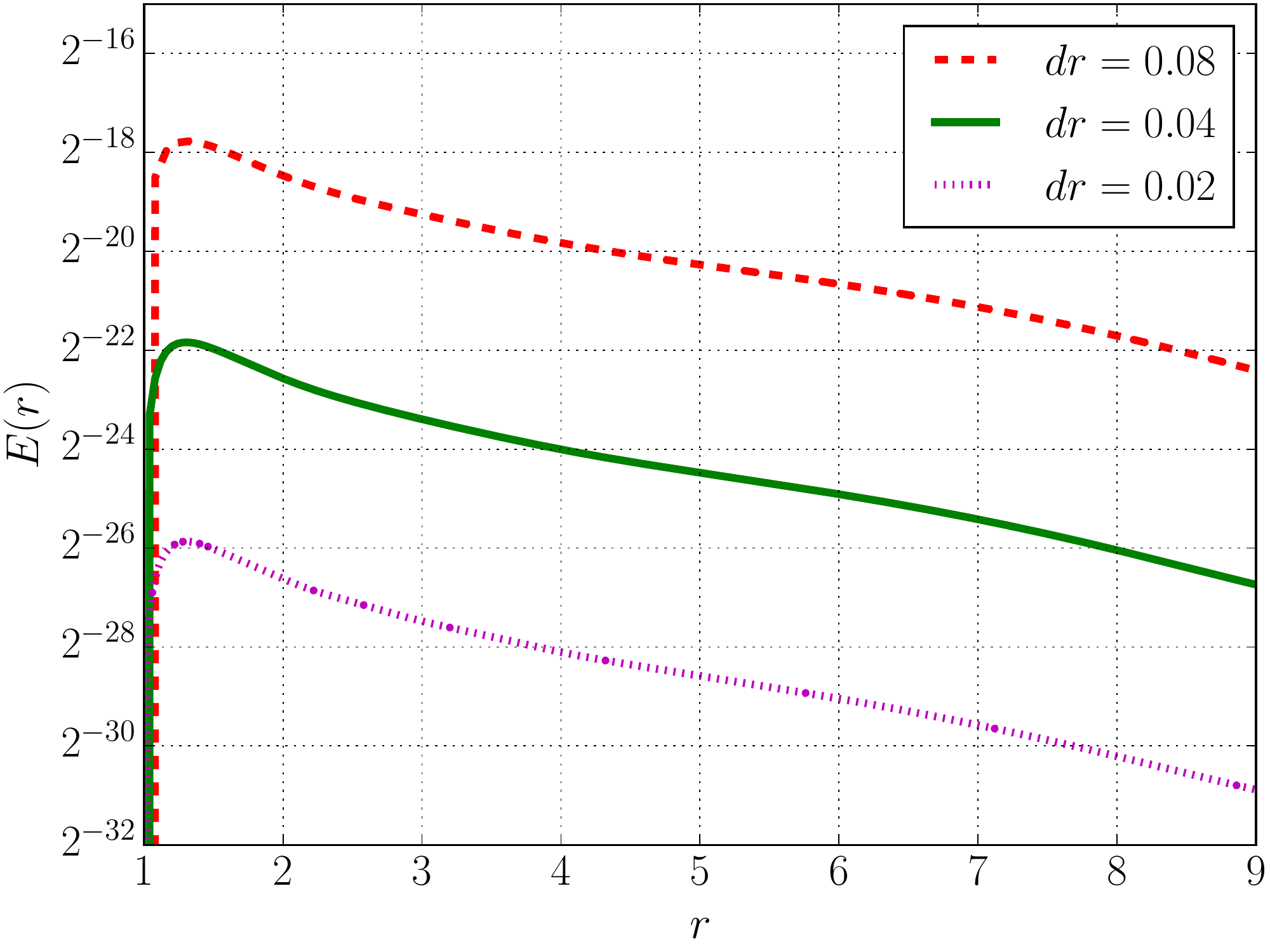}  
     \caption{Convergence using $r_0=1$, $N=32$, RK4 and various values of $dr$.} \label{fig:Convergence}    
  \end{minipage}  
  \hfill
  \begin{minipage}{0.49\linewidth}
    \centering
      \includegraphics[width=\linewidth]{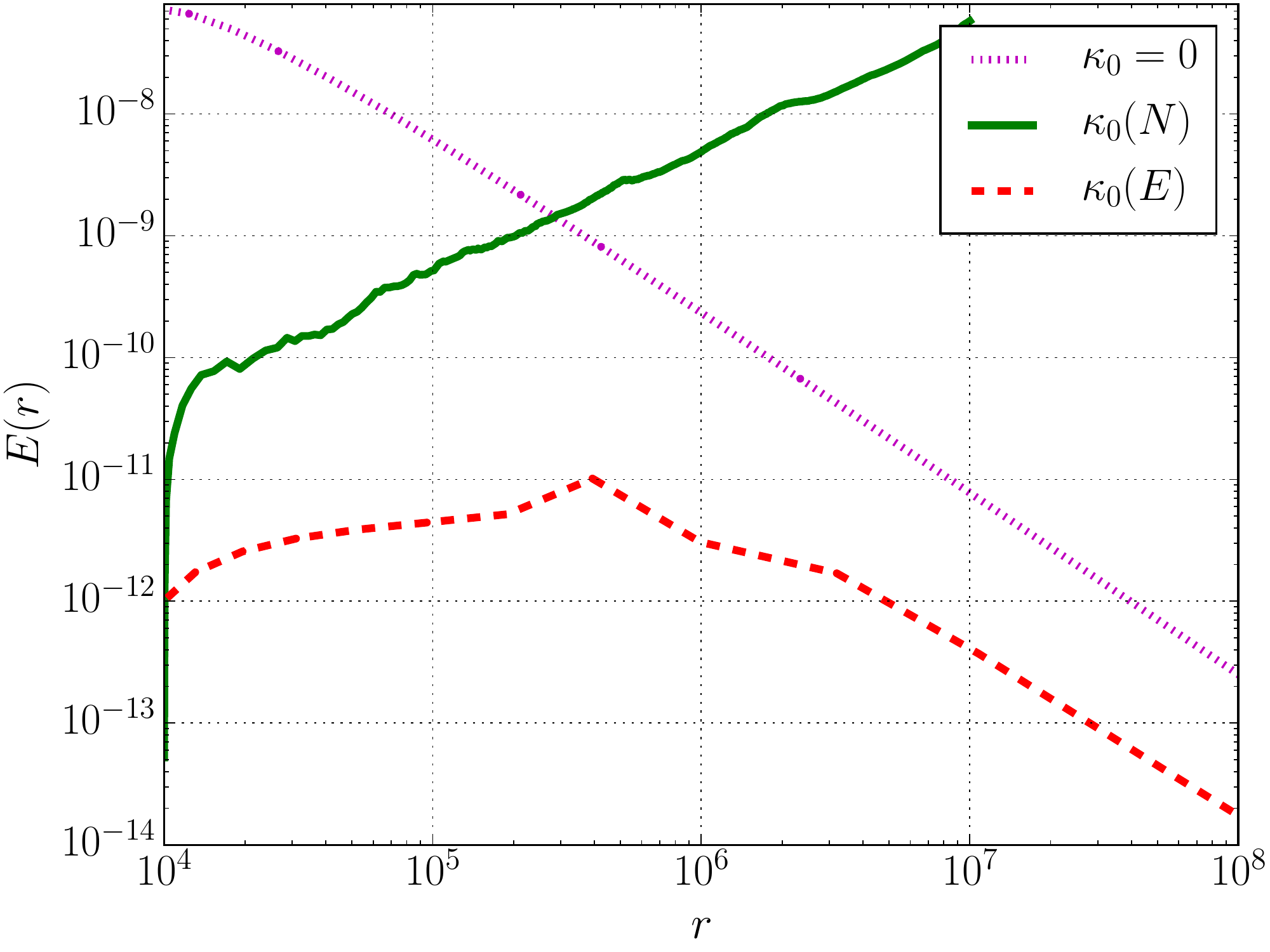}   
     \caption{Three different ways of numerically solving the equations (see the text) using $N=16$ and RKF (absolute tolerance $10^{-12}$).} \label{fig:numerical_problem_ko}
  \end{minipage}      
\end{figure}

In \Figref{fig:Convergence} we carry out a convergence test with RK4 (the 4th-order Runge-Kutta scheme) as the numerical integrator and $dr$ as the size of the ``time'' step. We see clearly that the numerical solutions converge in $4$th order which suggests that the numerical error is dominated by the ``time'' discretization. The ``spatial'' discretization error is expected to be negligible in this regime because the spectral discretization in space is expected to be highly accurate. All this is in full agreement with other numerical evolutions we have performed with the same code in previous investigations \cite{Beyer:2016fc,Beyer:2017jw}.

Notice that \Figref{fig:Convergence} restricts to small values of $r$. Depending on how we implemented the equations practically in our code, 
we were surprised to find the following numerical instability in our attempts to extend the evolutions to large $r$ values. We first attributed this to 
some sort of aliasing instability caused by non-linear terms in the equations. However, after a judicious study, we recognized terms of
the form $\kappa_0 / \BbbK$ in the equation as the source. In fact both quantities $\kappa_0$ and $\BbbK$ decay in the limit $r\rightarrow\infty$; \Eqref{def:KoInSchwarchild} yields the decay rate of $r^{-4}$ for $\kappa_0$ while  $\BbbK$ decays like $r^{-2}$ in the case of exact Kerr-Schild Kerr data according to  \Eqref{eq:BbbKKerr}. Later we will see that $\BbbK$ may decay slower than that for more general classes of solutions. But within the particular family of solutions considered in this paper it will always approach zero at infinity. As a consequence of the finite accuracy of numbers in computers, any expression which involves $\kappa_0 / \BbbK$ is therefore potentially problematic.

In \Figref{fig:numerical_problem_ko} we investigate this issue and demonstrate that the numerical error depends strongly on the way we numerically evaluate the $\kappa_0 / \BbbK$-terms.  We perform three different runs each of which has a different way to estimate these problematic terms. All these runs were performed with the Runge-Kutta-Fehlberg (RKF) method as the numerical integrator with absolute tolerance $10^{-12}$ and $N=16$.  
Recall that in general $\kappa_0$ is given by \Eqref{eq:constr3} which, in the case of Kerr-Schild Kerr data leads to an exact expression 
which is too long to be printed out here. Since we designed our code to be as general as possible, we wrote a routine that calculates the 
term ${ }^{(3)} R$ in \Eqref{eq:constr3} from any given free data $\hat\gamma_{ij}$, $\hat N$ and $\hat N^i$ fully numerically. While this
makes it in principle possible to perform runs with arbitrary free data without changing the code, it turns out that the associated numerical
errors are the source of the above mentioned instability. In fact, this routine is used in the run which yields the continuous (second) curve
in \Figref{fig:numerical_problem_ko}, for which the numerical error is clearly unstable. If we instead evaluate the terms $\kappa_0 / \BbbK$ 
in the equations by calculating ${ }^{(3)} R$ with exact computer algebra for Kerr-Schild Kerr data the numerical instability disappears as 
demonstrated by the dashed (third) curve  in \Figref{fig:numerical_problem_ko}. In comparison, we also perform a numerical run where we set 
$\kappa_0 / \BbbK$ as identically zero which should be a good approximation for large $r$ according to the decay rates given above. 
The associated numerical error yields the dotted (first) curve in \Figref{fig:numerical_problem_ko}.

%\textbf{TASK: Update these paragraphs.}
%These numerical tests clearly demonstrate that terms of the form $\kappa_0 / \BbbK$ are problematic numerically, and the way we calculate them can have a significant effect.
These tests clearly show that terms of the form $\kappa_0 / \BbbK$ may be problematic numerically if they are not treated with care. We have found, however, that this problem disappears when we pick the method corresponding
to the third curve in \Figref{fig:numerical_problem_ko}. This is what we will do in all of what follows. As a matter of fact, there are indeed several standard techniques in the literature for dealing with such issues which we could have employed as well (for instance, we could have rescaled the unknowns with appropriate powers of $r$). Without going into the details, we refer the interested reader to \cite{georgescu1995asymptotic}. 
%treating this short of problematic terms, however we have exclude them 
%from this work for the sake of simplicity. For the interested reader, however, can see for instance \cite{georgescu1995asymptotic}. 

We conclude this section with a quick remark about the CFL condition. Given that we use \textit{explicit} integrators, in particular the Runge-Kutta-Fehlberg method, one could think that the CFL condition may be violated in our numerical runs, for example for large values of $r$. If this was the case, our conclusions about the decay of the unknown quantities in the limit $r\rightarrow\infty$ below could not be trusted. Observe that while for  other codes, $r$ can range from small to  large values at \emph{each} time step, in particular when large spherical grids are used, it plays the role of the ``time parameter'' for us. In our case, the CFL condition therefore restricts the ``time'' step depending on the current \emph{single} value of $r$. Since the  Runge-Kutta-Fehlberg method adapts the time step dynamically during the evolution according to certain prescribed tolerances and truncation errors and the time step is therefore always compatible with the stability region of the integrator,  the CFL condition is always satisfied -- even for large values of $r$. For an extended discussion of related issues, see for instance \cite{burden2001numerical}.

% Before concluding this section we want to point out that because our pseudo-spectral approach allow us to compute exactly the angular derivatives by applying the 
% spin weight spherical harmonic transform, our PDE system can be seen as an ODE system with the radius r as a free parameter. As mentioned, for solving 
% this equivalent system, we use an adaptive radial-step explicit Runge Kutta integrator (Runge-Kutta-Fehlberg) method. Rather than determining the radial-step by invoking 
% the CFL condition (using a fixed CFL parameter determined a priori), in this scheme the size of the radial-interval is chosen dynamically such that the integrator is always driven
% into its region of stability. This size is fixed by some given tolerance and a certain truncation error. When this error is bigger than the tolerance, the 
% radial step is reduced by a certain factor. Otherwise, the algorithm increases the step size in order to make the integration process more efficient 
% (for an extended discussion on this method see, for instance, \cite{burden2001numerical}). Therefore the CFL condition no longer limits the size of the radial-step.

\section{Asymptotics of nonlinearly perturbed Schwarzschild initial data}
\label{sec:analysisasymptotics}

\subsection{Spherically symmetric data}
\label{sec:sphericallysymmdata}

Now we want to present the main results of this paper in detail. Let us start by recalling the family of exact
  solutions of the constraint equations given by \Eqsref{schwarzschildvaluesinitiadata} and \eqref{ec:schwarchildcase} for any $C\ge 0$. The exact Kerr-Schild Schwarzschild initial data set corresponds to $C=0$. The first main question we address is whether all members of this family of spherically symmetric initial data sets are asymptotically flat.

The metric $h_{ij}$ given by the free data \Eqref{schwarzschildvaluesinitiadata} takes the form
\begin{equation}
  \label{eq:SphSymIntrinsicMetric}
  h_{ij}=\left(1+\frac{2M}r\right) dr^2+r^2 d\theta^2+r^2\sin^2\theta d\varphi^2.
\end{equation}
When we introduce the new radial coordinate 
\begin{equation}
  \label{eq:newradialcoordinate}
  R=r-M+\frac{3M^2}{4r},
\end{equation}
it takes the form 
\[h_{ij}=\left(1+\frac{2M}{R}+O(R^{-2})\right)\left(dR^2+R^2 d\theta^2+R^2\sin^2\theta d\varphi^2\right).\]
The metric is therefore consistent with both the weak and the strong notions of asymptotic flatness when $R$ is identified with $\varrho$ (and when Cartesian coordinates are introduced in the standard way from polar coordinates $(R,\theta,\varphi)$), cf.\ \Eqsref{eq:AFLimit}, \eqref{eq:AFStrong1} and \eqref{eq:AFWeak1}.

Regarding the second fundamental form, we find
\begin{equation}
  \label{eq:chiintermsofdata}
  \chi:=h^{ij}\chi_{ij}=\text{\boldmath$\kappa$}+\BbbK,
\end{equation}
which follows directly from \Eqref{ec:mean_curvature_decomposition}.
A necessary condition for the validity of the limit \Eqref{eq:AFStrong2} (or \eqref{eq:AFWeak2}) is therefore that
$\chi=O(R^{-2})$ (or $\chi=o(R^{-3/2})$) in the limit $R\rightarrow \infty$, which is equivalent to $\chi=O(r^{-2})$ (or $\chi=o(r^{-3/2})$) in the limit $r\rightarrow \infty$. Given that
\begin{equation}  
  {\BbbK}=
  \begin{cases}
    O(r^{-2}) & \text{if $C=0$},\\
     O(r^{-3/2}) & \text{if $C>0$},
  \end{cases}
\end{equation}
as a consequence of \Eqref{ec:schwarchildcase} 
it follows that
\begin{equation}
  \chi=
  \begin{cases}
    O(r^{-2}) & \text{if $C=0$},\\
     O(r^{-3/2}) & \text{if $C>0$},
  \end{cases}
\end{equation}
using that
$\text{\boldmath$\kappa$}$ can be calculated from \Eqref{eq:constr3} and the quantities before. 
In conclusion, if $C>0$, then \emph{both} limits \Eqsref{eq:AFStrong2} and \eqref{eq:AFWeak2} are violated. The initial data set is therefore not asymptotically flat, neither in the strong 
nor in the weak sense. Therefore, only in the case $C=0$, the initial data set is asymptotically flat (both in the strong and the weak sense). In the latter case, one can that show that also all derivatives of these fields decay with appropriate rates. In particular all curvature quantities therefore decay at infinity. This is the case only for $C=0$.

% Note that 
% both notions of asymptotic flatness only depends on the fall off of the metric and the extrinsic curvature ignoring the behavior of
% other important quantities, such as the curvature tensor or the Ricci scalar.

% \textbf{TASK: Move this last sentence to the introduction and clarify it.}

\begin{figure}[t]
  \centering
  \includegraphics[width=0.6\linewidth]{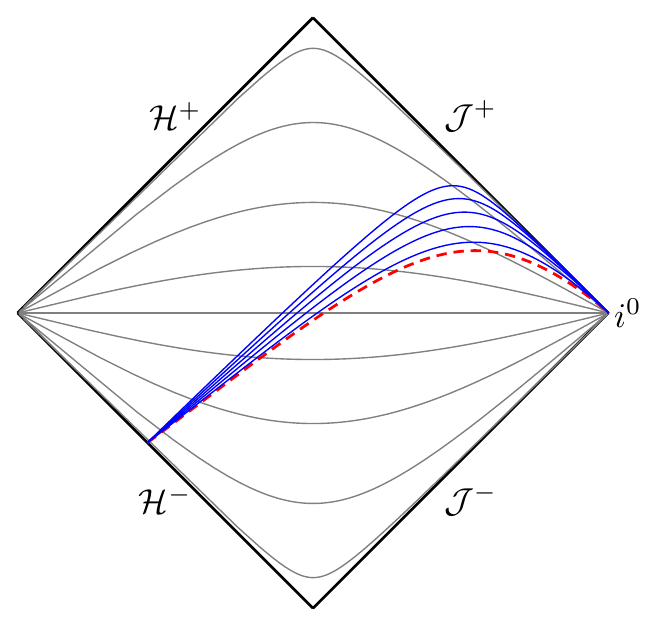}  
  \caption{Isometric embeddings of spherically symmetric initial data slices into the domain of outer communications of the Schwarzschild spacetime with mass $m=1$. Gray curves represent surfaces of constant Schwarzschild time. The (red) dashed curve is a surface corresponding to $C=0$, $M=m=1$ and the negative sign in \Eqref{ec:schwarchildcase} (i.e., a surface of constant Kerr-Schild time). The (blue) solid curves represent surfaces for different values of $C>0$ according to \Eqref{eq:CfromMassDiff} for $M=0.8, 0.4, 0, -0.4, -0.8$ (i.e., $M\not=m$) and the negative sign in \Eqref{ec:schwarchildcase}. Observe that the case $M=0.8$ is the curve closest to the $C=0$ curve before. Null infinity is represented by $\mathcal J^\pm$, spacelike infinity by $i^0$ and the horizon by $\mathcal H^{\pm}$. Notice that if $M<-m$, the corresponding surface is not spacelike close the horizon. Surfaces corresponding to the positive sign in \Eqref{ec:schwarchildcase} are obtained by flipping the above diagram upside down. } \label{fig:SurfacesSS}    
 \end{figure}

Owing to the fact that the two cases $C=0$ and $C>0$ are therefore significantly different we shall spend the remainder of this subsection on the following geometric interpretation of these initial data sets. We first observe that since these initial data sets are all spherically symmetric, it must be possible to (locally) embed them isometrically into the Schwarzschild or Minkowski spacetime as a consequence of Birkhoff's theorem. In fact, we can construct these isometric embeddings explicitly. They are depicted in \Figref{fig:SurfacesSS} and can be described as follows:

\newcommand{\SchwS}{\mathrm{SS}}
\begin{description}
\item [Case $C=0$:]
Let $(\Sigma, h_{ij})$ be the $3$-dimensional Riemannian manifold where the metric is given by \Eqref{eq:SphSymIntrinsicMetric} for any $M\ge 0$. As before, we use coordinates $(r,\theta,\varphi)$ on $\Sigma$.
The two possible families of isometric embeddings of $(\Sigma, h_{ij})$ into the Schwarzschild spacetime with mass $m=M$ are
\begin{equation}
\label{eq:KSembedding}
  t_{\SchwS}=t_0\pm 2 m \log (r-2 m),\quad r_{\SchwS}=r, \quad  \theta_{\SchwS}=\theta, \quad  \varphi_{\SchwS}=\varphi,
\end{equation}
for any $t_0\in\R$ and $r>2M$, where $(t_{\SchwS}, r_{\SchwS}, \theta_{\SchwS}, \varphi_{\SchwS})$ are standard Schwarzschild coordinates. The pull-back of the second fundamental form induced from the Schwarzschild metric with mass $m$ on any such slice agrees with the field $\chi_{ij}$  determined as in \Sectionref{sec:Setup} from the fields given by \Eqsref{schwarzschildvaluesinitiadata}, \eqref{def:KoInSchwarchild} and \eqref{ec:schwarchildcase} for $C=0$, $M=m$ and the sign $\mp$ in \Eqref{ec:schwarchildcase}. Notice that \Eqref{eq:KSembedding} is the well-known formula for (in- and outgoing) Kerr-Schild constant-time slices with respect to standard  Schwarzschild coordinates. 
\item [Case $C>0$:] 
Let $(\Sigma, h_{ij})$ be the $3$-dimensional Riemannian manifold where the metric is given by \Eqref{eq:SphSymIntrinsicMetric} for any $M\in\R$. As before, we use coordinates $(r,\theta,\varphi)$ on $\Sigma$.
The two possible families of isometric embeddings of $(\Sigma, h_{ij})$ into \emph{any} Schwarzschild spacetime with mass $m>\max\{0,M\}$ (i.e., $m\not= M$) are
\begin{align*}
  t_{\SchwS}&=t_0\pm\Biggl[m \log \frac{m \left(2 M+r -2 \sqrt{4 M m-2 M r+2 m r}\right)-M r+2 m^2}{m
   \left(2 M+r +2 \sqrt{4 M m-2 M r+2 m r}\right)-M r+2 m^2}\\
  &\qquad\qquad+2 \sqrt{4 M
   m-2 M r+2 m r}-4m+2m\log\frac{8m^2}{m-M}\Biggr],\\
  r_{\SchwS}&=r, \quad  \theta_{\SchwS}=\theta, \quad  \varphi_{\SchwS}=\varphi,
\end{align*}
for any $t_0\in\R$ and $r>2\max\{m,-M\}$, where, as above, $(t_{\SchwS}, r_{\SchwS}, \theta_{\SchwS}, \varphi_{\SchwS})$ are standard Schwarzschild coordinates. The pull-back of the second fundamental form induced from the Schwarzschild metric with mass $m$ on any such slice agrees with the field $\chi_{ij}$  determined as in \Sectionref{sec:Setup} from the fields given by \Eqsref{schwarzschildvaluesinitiadata}, \eqref{def:KoInSchwarchild} and \eqref{ec:schwarchildcase} for 
\begin{equation}
  \label{eq:CfromMassDiff}
  C=8 (m - M),
\end{equation}
and the sign $\mp$ in \Eqref{ec:schwarchildcase}. Observe that this finding is in full consistency with a result by \'O Murchadha and Roszkowski in \cite{Murchadha:2005ib} (even though the authors do not provide the explicit embedding formulas). 
\end{description}

Loosely speaking, the case $C>0$ yields initial data sets with the ``wrong mass'' $m\not= M$; it is therefore not surprising that such initial data sets are not asymptotically flat. We have found that within the family of spherically symmetric slices in the Schwarzschild spacetime, this subfamily of slices with the ``wrong mass'' is generic. The main practical consequence for the initial value problem of the hyperbolic formulation of the constraint equations in \Sectionref{sec:hyperbolicconstraints} is that the \emph{initial data must therefore be fine-tuned in order to obtain an asymptotically flat initial data set} (i.e., one with the ``right mass'' $m=M$). While in the spherically symmetric case, we know explicitly how to do this, the problem of finding asymptotically flat initial data sets in less symmetric situations using the approach in \Sectionref{sec:hyperbolicconstraints}  is therefore likely to be very hard. In order to obtain further insights
we shall now study perturbations of the above data sets in the axially symmetric case.

\subsection{Axially symmetric data}
\label{sec:heuristics}

In this subsection we keep the ``exact Kerr-Schild Schwarzschild free data'' in \Eqref{schwarzschildvaluesinitiadata}, but then allow in principle arbitrary axially symmetric solutions $\BbbK$ and $\Bbbkm$ of the fully nonlinear system \Eqref{sys:simplifiedsystemofequations} (as long as the hyperbolicity condition \Eqref{eq:hyperbolicitycondition} holds). The resulting initial data sets are therefore axially symmetric nonlinear perturbations of the Kerr-Schild Schwarzschild initial data set in the sense discussed in \Sectionref{sec:exactsols}. 
The objective is then  to  study the asymptotics of these data sets in the limit $r\rightarrow\infty$ in the light of our results of the spherically symmetric case in \Sectionref{sec:sphericallysymmdata}.
In a first step we restrict to heuristic and approximate studies in order to develop a first feeling of what might happen. In \Sectionref{sec:numericalsupport} we shall then tackle the fully nonlinear setting numerically.

To this end, we suppose that we have a one-parameter family of axially symmetric solutions of \Eqref{sys:simplifiedsystemofequations} of the form
\begin{equation}
  \label{eq:linansatz}
  \BbbK = \BbbK^{(0)} + \epsilon\, \BbbK^{(1)},\quad
  \Bbbkm=\epsilon\, \Bbbkm^{(1)},
\end{equation}
where $\BbbK^{(0)}$ only depends on $r$ and where $\epsilon$ is a small parameter.
When we plug this ansatz into \Eqref{sys:simplifiedsystemofequations} and evaluate it for $\epsilon=0$, the  equations become that of the spherically symmetric case \Eqref{eq:appriximation3}, i.e.,
\begin{equation*}
\partial_r \BbbK^{(0)} = -\frac{3 \BbbK^{(0)} }{2 r} -\frac{\kappa_0}{r \BbbK^{(0)} }, 
\end{equation*}
for $\kappa_0$ given by \Eqref{def:KoInSchwarchild}. As before, the general solution, which is real for all large values of $r$, is 
\begin{equation}\label{ec:schwarchildcase2}
  \BbbK^{(0)} =  \pm \dfrac{1}{r^{3/2}} \sqrt{ C + \frac{16 M^2}{r+2 M} },
\end{equation}
for any $C\ge 0$. 

Next we \emph{average} the first equation in \Eqref{sys:simplifiedsystemofequations} over any surface $S_r$ with $r>0$, i.e., we integrate with respect to $\theta$ (and $\varphi$, but recall that we restrict to the axially symmetric case) using the standard volume element of the round unit $2$-sphere and then normalize this by dividing by $4\pi$. When we now (i) use the fact that such an average of the term $\eth\Bbbkmbar  + \bar{\eth }\Bbbkm$ in this equation is zero (because $\Bbbkm$ and $\Bbbkmbar$ have spin-weight $1$ and $-1$, respectively), (ii) consider the term $4 |\Bbbkm|^2/(r^3 \BbbK) $ in the equation as negligible (because it is $O(\epsilon^2)$), and, (iii) observe that
\[\frac 1{\BbbK}=\frac 1{\BbbK^{(0)}}-\epsilon\frac {\BbbK^{(1)}}{(\BbbK^{(0)})^2}+O(\epsilon^2),\]
we find that we can identify $\BbbK^{(0)}(r)$ with the average 
\begin{equation}
  \label{eq:averageK}
  \underline{\BbbK}(r)=\frac 1{4\pi}\int_0^{2\pi}\int_{0}^\pi \BbbK(r,\theta,\varphi) \sin\theta d\theta d\varphi
\end{equation}
in leading-order in $\epsilon$.
The average of $\BbbK^{(1)}$ in \Eqref{eq:linansatz} must therefore be identically zero at every $r$.
Because of this we shall now always write $\underline{\BbbK}$ instead of $\BbbK^{(0)}$ in all of what follows. From \Eqref{ec:schwarchildcase2}, it thus follows that
\begin{equation}
  \label{eq:asymptotics1}
  \underline{\BbbK}=
  \begin{cases}
    O(r^{-2}) & \text{if $C=0$},\\
     O(r^{-3/2}) & \text{if $C>0$},
  \end{cases}
\end{equation}
in the limit $r\rightarrow\infty$. 

Observe that in the language of spin-weighted spherical harmonics, see \Sectionref{sec:spheresandeth}, $\underline{\BbbK}=\BbbK^{(0)}$ is proportional to the $l=0$-mode of $\BbbK$, while $\BbbK^{(1)}$ is given by the collective contributions of all $l>0$-modes and hence describes the variations of $\BbbK$ along the spheres at each value of $r$.

Now let us derive certain bounds for $\BbbK^{(1)}$ and $\Bbbkm$ according to the ansatz \Eqref{eq:linansatz} which shall consequently allow us to deduce the asymptotics of these quantities. To this end, we \emph{assume} that $\epsilon$ is so small such that
\[|\epsilon\, \BbbK^{(1)}|<\BbbK^{(0)}\] 
holds for every $r$, $\theta$ and $\varphi$. If this is true it implies that $|\BbbK|<2|\underline{\BbbK}|$ everywhere. We can thus conclude that $\BbbK$ decays (at least) as fast as $\underline{\BbbK}$ in \Eqref{eq:asymptotics1}.
If we also \emph{assume}  that the hyperbolicity condition \eqref{eq:hyperbolicitycondition} holds everywhere, we can deduce that
\[0>2 \mathbf{k}^i \mathbf{k}_i - \dfrac{  \mathbf{K} ^2 }{2}  - \kappa_0 
>4 r^{-2} |\Bbbkm|^2 -2{  \underline{\BbbK} ^2 }  - \kappa_0
\quad\Longleftrightarrow\quad  |\Bbbkm|^2<\frac{r^2}4\left(2{  \underline{\BbbK} ^2 }  + \kappa_0\right)\]
where we used \Eqref{eq:constr3} and previous estimates.
Since $\kappa_0 = {O}(r^{-4})$ (see \Eqref{def:KoInSchwarchild}), \Eqref{eq:asymptotics1} implies
\begin{equation}
  \label{eq:asymptotics2}
  \Bbbkm=
  \begin{cases}
    O(r^{-1}) & \text{if $C=0$},\\
     O(r^{-1/2}) & \text{if $C>0$}.
  \end{cases}
\end{equation}
Notice that this may not at all be a sharp decay estimate. In fact, \emph{the better the hyperbolicity condition is satisfied, i.e., the more negative the function $\text{\boldmath$\kappa$} \ \mathbf{K}$ is, the less optimal we expect \Eqref{eq:asymptotics2} to be}.

 In any case, given all these estimates, \Eqref{eq:constr3} now implies that
\begin{equation}
  \text{\boldmath$\kappa$}=
  \begin{cases}
    O(r^{-2}) & \text{if $C=0$},\\
     O(r^{-3/2}) & \text{if $C>0$},
  \end{cases}
\end{equation}
and \Eqref{eq:chiintermsofdata} yields
\begin{equation}
  \label{eq:asymptoticschi}
  \chi=
  \begin{cases}
    O(r^{-2}) & \text{if $C=0$},\\
     O(r^{-3/2}) & \text{if $C>0$}.
  \end{cases}
\end{equation}

Since the free data, and therefore the metric $h_{ij}$, are the same as in the spherically symmetric case in \Sectionref{sec:sphericallysymmdata}, the same analysis regarding asymptotic flatness in \Sectionref{sec:sphericallysymmdata} can be performed here. It  suggests that generic resulting initial data sets (i.e., those corresponding to $C>0$) are not asymptotically flat in the strong sense, see in particular \Eqref{eq:AFStrong2}. Regarding the weak notion of asymptotic flatness, see in particular \Eqref{eq:AFWeak2}, the situation is more subtle. In fact, if the decay exponent of $\chi$ is $3/2+\epsilon$ for an arbitrarily small $\epsilon>0$, the initial data set may be asymptotically flat in this weak sense. Arguably it appears to be unlikely  that the second fundamental form of perturbed data sets, which we consider here, would decay better than the second fundamental form of generic unperturbed (spherically symmetric) data sets given by $C>0$. In any case, the next subsection is devoted to the fully nonlinear study of such issues.

Before we do this however let us close this subsection with the following remark regarding the case $C=0$ for which our analysis above suggests asymptotic flatness (both in the weak and the strong sense). 
Recall that our analysis here was based on neglecting terms of higher order in $\epsilon$ and therefore on neglecting nonlinear couplings between the $l=0$-mode and all modes $l>0$. As a consequence of this the average $\underline{\BbbK}$ was a solution of the ``spherically symmetric equations'' which have the exact solution \Eqref{ec:schwarchildcase2}. Nonlinear interactions of the modes may however produce significant deviations in particular in the limit $r\rightarrow\infty$. It is therefore conceivable that,   even if we pick initial data for the hyperbolic constraints which correspond to $C=0$, the asymptotics will nevertheless be described by the \emph{generic} case $C>0$.

\subsection{Numerical analysis}
\label{sec:numericalsupport}
The purpose of this section is to support the largely heuristic arguments presented in the previous subsection by numerically solving the fully nonlinear system of constraints. This will allow us to confirm the decay rates derived before and thereby to provide support for our claims regarding asymptotic flatness of the resulting initial data sets. To this end we choose the same free data in \Eqref{schwarzschildvaluesinitiadata} as before, but now we solve the fully nonlinear system \Eqref{sys:simplifiedsystemofequations} numerically. As the initial hypersurface we pick $S_{r_0}$ with $r_0 = 10^{3}$ and as initial data we pick the family
\begin{eqnarray*}
\BbbK|_{r=r_0} &=&  -\dfrac{1}{r_0^{3/2}} \frac{4 M}{\sqrt{ r_0+2 M} } - \beta \ {}_{0}Y_{0} (\theta)  - \epsilon \  {}_{0}Y_{2} (\theta) , \\
\Bbbkm  |_{r=r_0} &=& \epsilon  \ {}_{1}Y_{1} (\theta),
\end{eqnarray*}
for parameters $\beta\ge 0$ and $\epsilon\in\R$ which implies that $\underline{\BbbK}|_{r=r_0}$ is given by \Eqref{ec:schwarchildcase2} with $r=r_0$ and
\begin{equation}
  \label{eq:relCbeta}
  C=16 M^2\left(\frac{4 M r_0^{3/2}}{\sqrt{\pi } \sqrt{2 M+{r_0}}}+\frac{\beta r_0^3}{4 \pi }\right)\beta\ge 0.
\end{equation}
In the following we always make sure that $\epsilon$ is sufficiently  small in comparison to $\beta$ so that $\BbbK|_{r=r_0}$ never vanishes anywhere.
Observe that we use the ``axial symmetric notation'' here introduced in \cite{Beyer:2016fc}, that is,  ${ Y_{l}}(\theta) = { {}_{0} Y_{l0}}(\theta,\varphi)$ and ${ {}_{1} Y_{l}}(\theta) = { {}_{1}  Y_{l0}}(\theta,\varphi)$.  Moreover, we write
\begin{equation}
  \label{eq:modedecomposition}
  \BbbK(r,\theta)=\sum_{l=0}^\infty a_l(r) { Y_{l}}(\theta),\quad \Bbbkm (r,\theta)=\sum_{l=1}^\infty b_l(r)\, {{}_{1} Y_{l}}(\theta).
\end{equation}
In all the numerical evolutions we choose $N=16$ as the number of $\theta$-grid points. The inequality
\eqref{eq:hyperbolicitycondition} is always satisfied during all numerical evolutions. Hence, the system
\eqref{sys:simplifiedsystemofequations} always remains hyperbolic. In the following plots we show the behavior of
the modes $a_l(r)$ and $b_l(r)$ up to order four. In all these figures we have picked constants $\alpha$ and $\tilde\alpha$
in order to place certain reference curves optimally. Furthermore, for a better appreciation of the different orders of
magnitude, we have used a logarithmic scale for the horizontal axis and a ``symmetric logarithmic scale'' for
the vertical axis (the ``symlog'' option in matplotlib in python, for details see for instance \cite{webber2012bi}).

\begin{figure}[t]
  \begin{minipage}{0.49\linewidth}
    \centering
     \includegraphics[width=\linewidth]{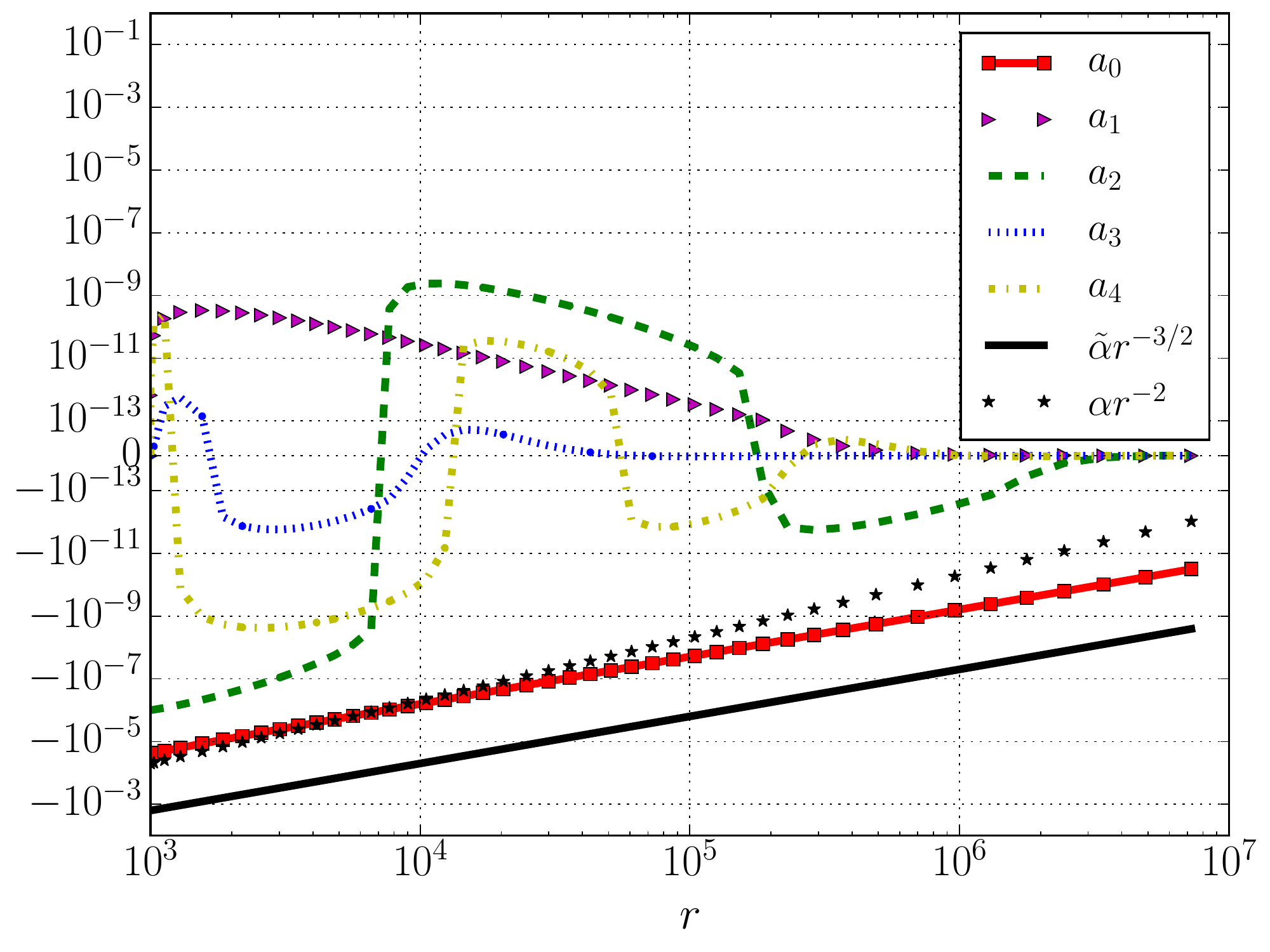}  
     \caption{Numerical evolutions and decay rates for the first modes of $\BbbK$ (Case~1).} \label{fig:KmodesN}    
  \end{minipage}  
  \hfill
  \begin{minipage}{0.49\linewidth}
    \centering
      \includegraphics[width=\linewidth]{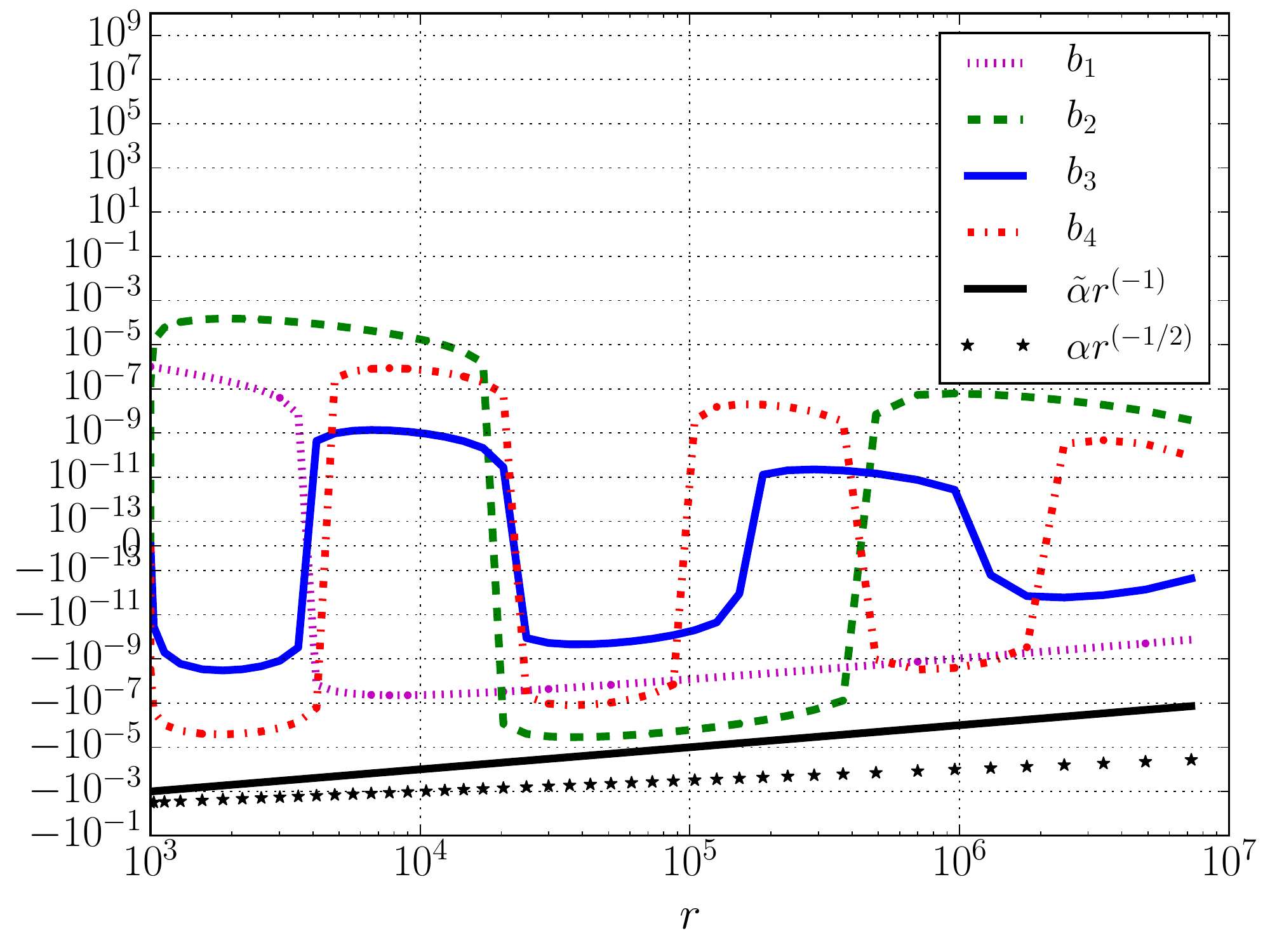}   
      \caption{Numerical evolutions and decay rates for the first modes of $\Bbbkm$ (Case~1).}
     \label{fig:klmodesN}
  \end{minipage}    
  
  \,\hfill\begin{minipage}{0.49\linewidth}
    \centering
      \includegraphics[width=\linewidth]{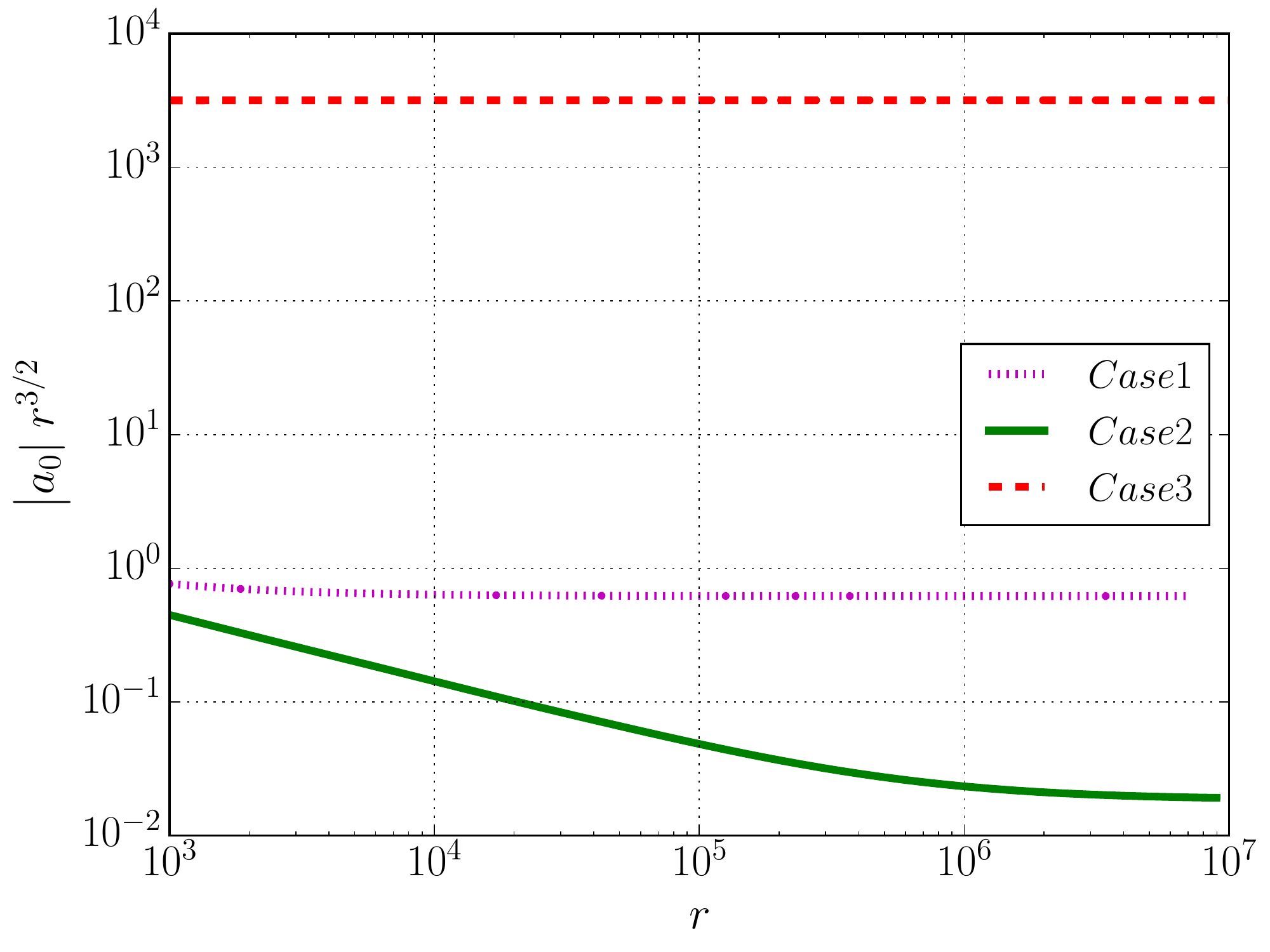}   
      \caption{Demonstration that $a_0$ decays like $r^{-3/2}$ asymptotically for all three cases considered in the text.}
     \label{fig:decaya0}
  \end{minipage}\hfill\,

\end{figure}

\paragraph{Case 1 ($\beta = 10^{-5}$ and $\epsilon = 10^{-6}$).}
According to \Eqref{eq:relCbeta} this corresponds to the case $C>0$. From the discussion in \Sectionref{sec:heuristics}, we expect that  $\BbbK={O}(r^{-3/2})$ and $\Bbbkm={O}(r^{-1/2})$. This is confirmed in \Figsref{fig:KmodesN} and \ref{fig:klmodesN}. In particular, we see that $a_0$ (recall \Eqref{eq:modedecomposition}) definitely decays slower than $r^{-2}$ and hence (as one can check) the strong notion of asymptotic flatness is violated. There is no evidence that it decays faster than $r^{-3/2}$ (which would be necessary for the weak notion). In fact, the dotted curve in \Figref{fig:decaya0}, which shows $r^{3/2} a_0$, suggests that $a_0$ decays exactly like $r^{-3/2}$, therefore violating the weak notion of asymptotic flatness.

\begin{figure}[t]
  \begin{minipage}{0.49\linewidth}
    \centering
     \includegraphics[width=\linewidth]{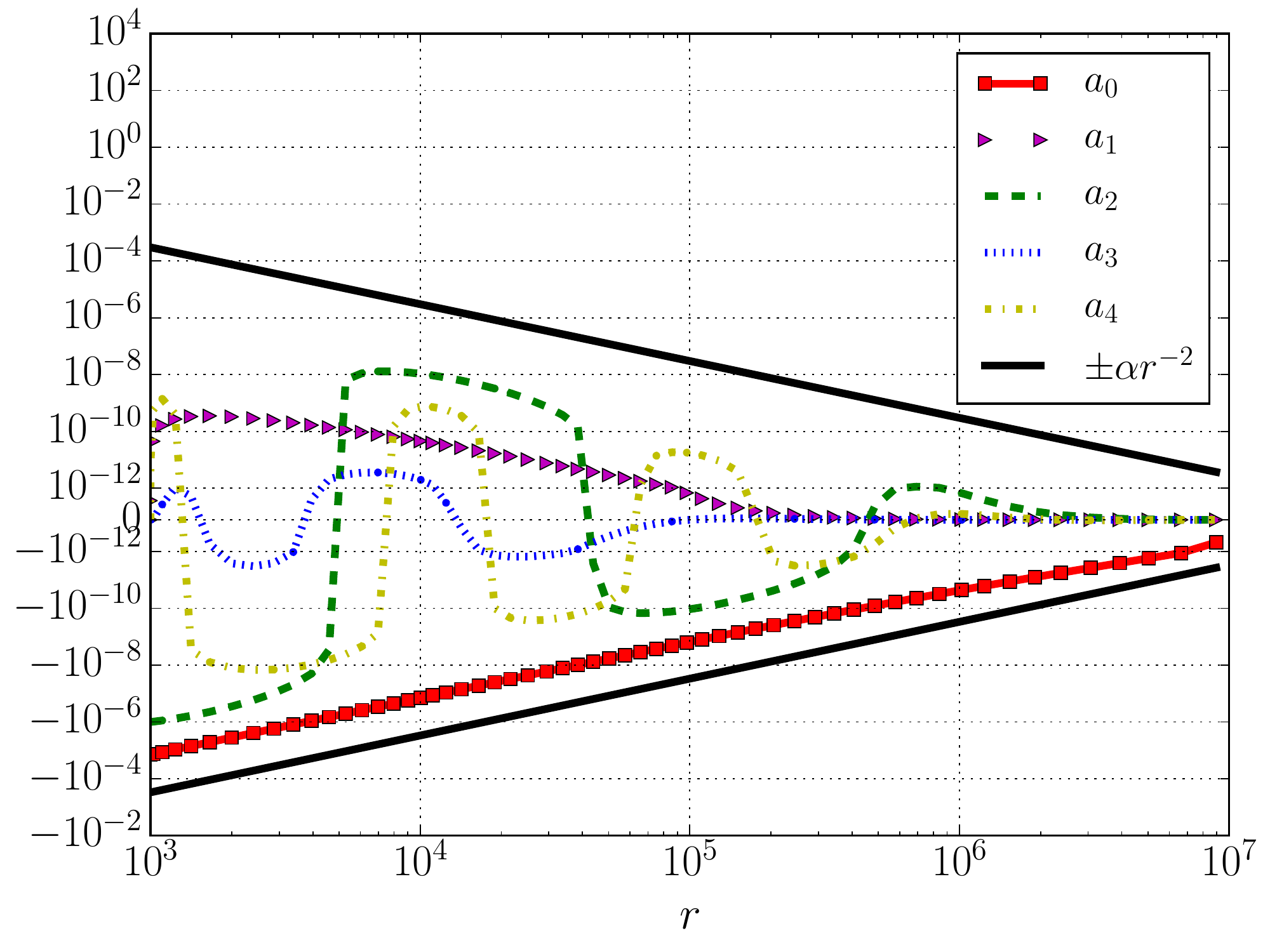}  
     \caption{Numerical evolutions and decay rates for the first modes of $\BbbK$ (Case~2).} \label{fig:KmodesS}    
  \end{minipage}  
  \hfill
  \begin{minipage}{0.49\linewidth}
    \centering
      \includegraphics[width=\linewidth]{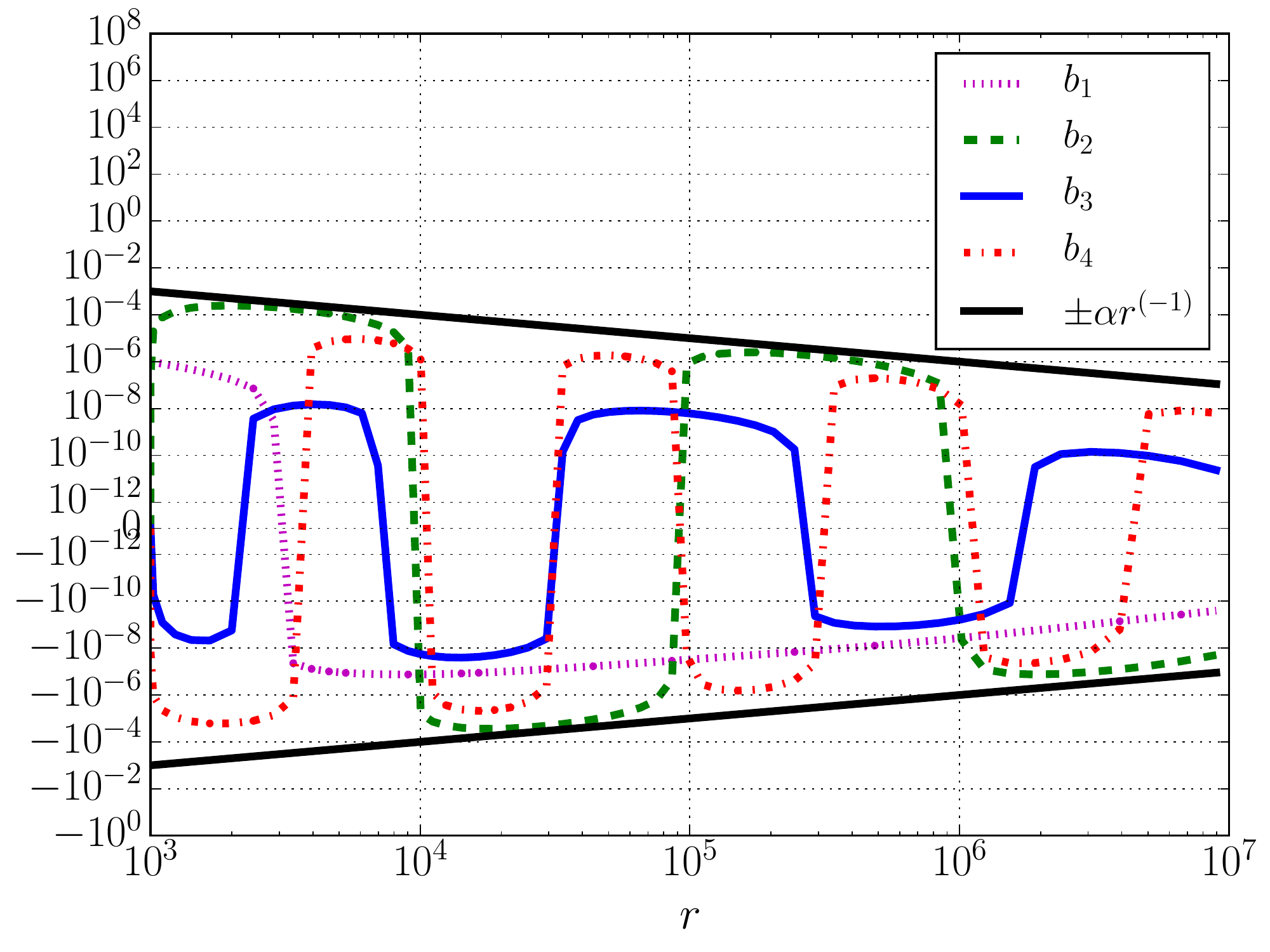}   
     \caption{Numerical evolutions and decay rates for the first modes of $\Bbbkm$ (Case~2).} \label{fig:klmodesS}
  \end{minipage}      
\end{figure}

\paragraph{Case~2 ($\beta = 0$ and $\epsilon = 10^{-6}$).} This is an example of the (presumably) unstable case corresponding to $C=0$. It is very interesting that the numerical results in \Figsref{fig:KmodesS} and \ref{fig:klmodesS} seem to be largely compatible with the decay rates in \Sectionref{sec:heuristics} despite our argument in the last paragraph there. The solid curve in \Figref{fig:decaya0}, however, clearly shows that while the decay rate of $a_0$ is faster for some period of the evolution (in fact, there, the decay is compatible with the predictions in \Sectionref{sec:heuristics}), eventually, the decay rate approaches $-3/2$ as expected.  The strong notion of asymptotic flatness is therefore definitely violated; most likely also the weak notion. 

\begin{figure}[t]
  \begin{minipage}{0.49\linewidth}
    \centering
     \includegraphics[scale=0.36]{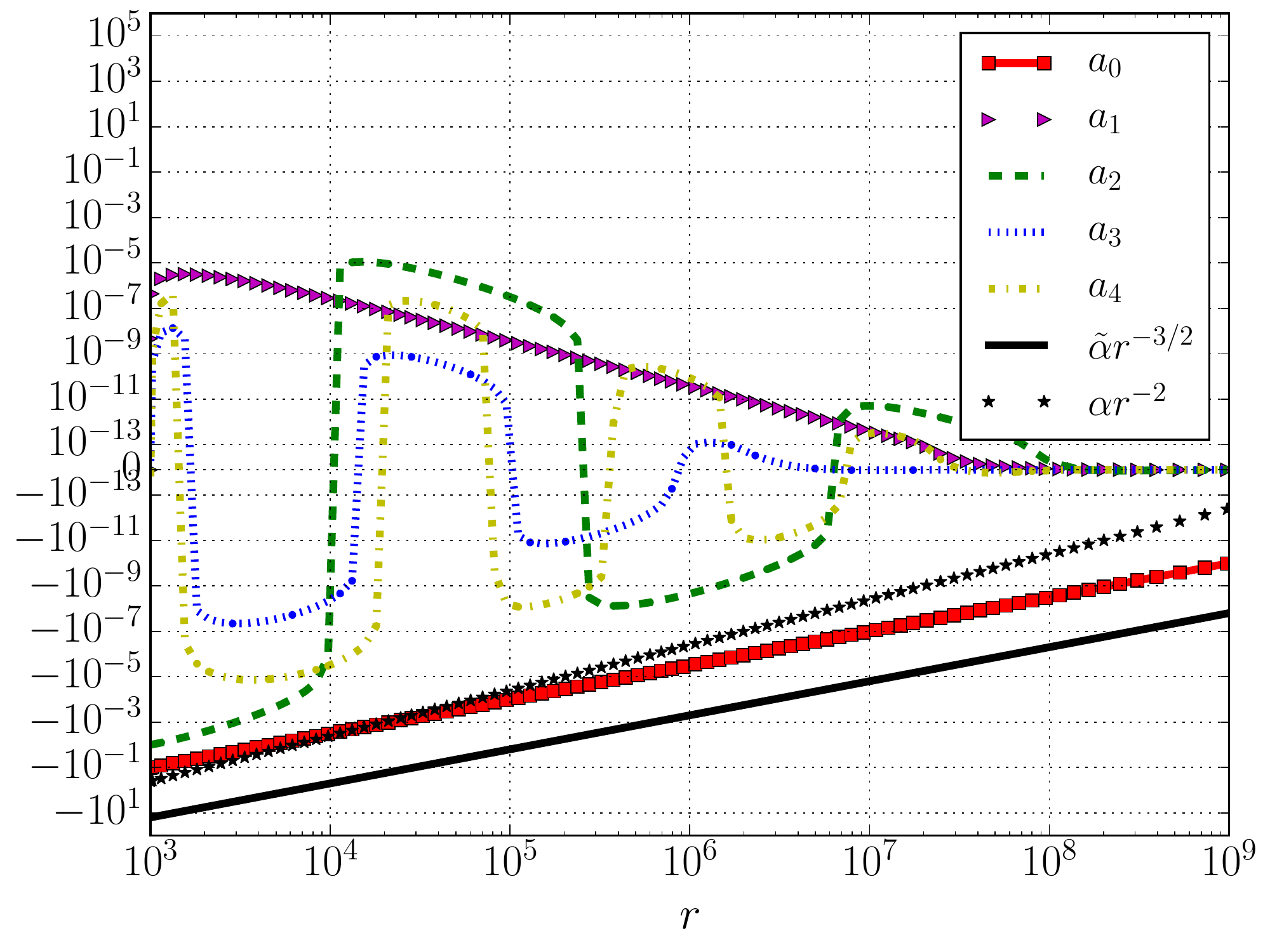}  
     \caption{Numerical evolutions and decay rates for the first modes of $\BbbK$ (Case~3).} \label{fig:KmodesNLARGE}    
  \end{minipage}  
  \hfill
  \begin{minipage}{0.49\linewidth}
    \centering
      \includegraphics[scale=0.36]{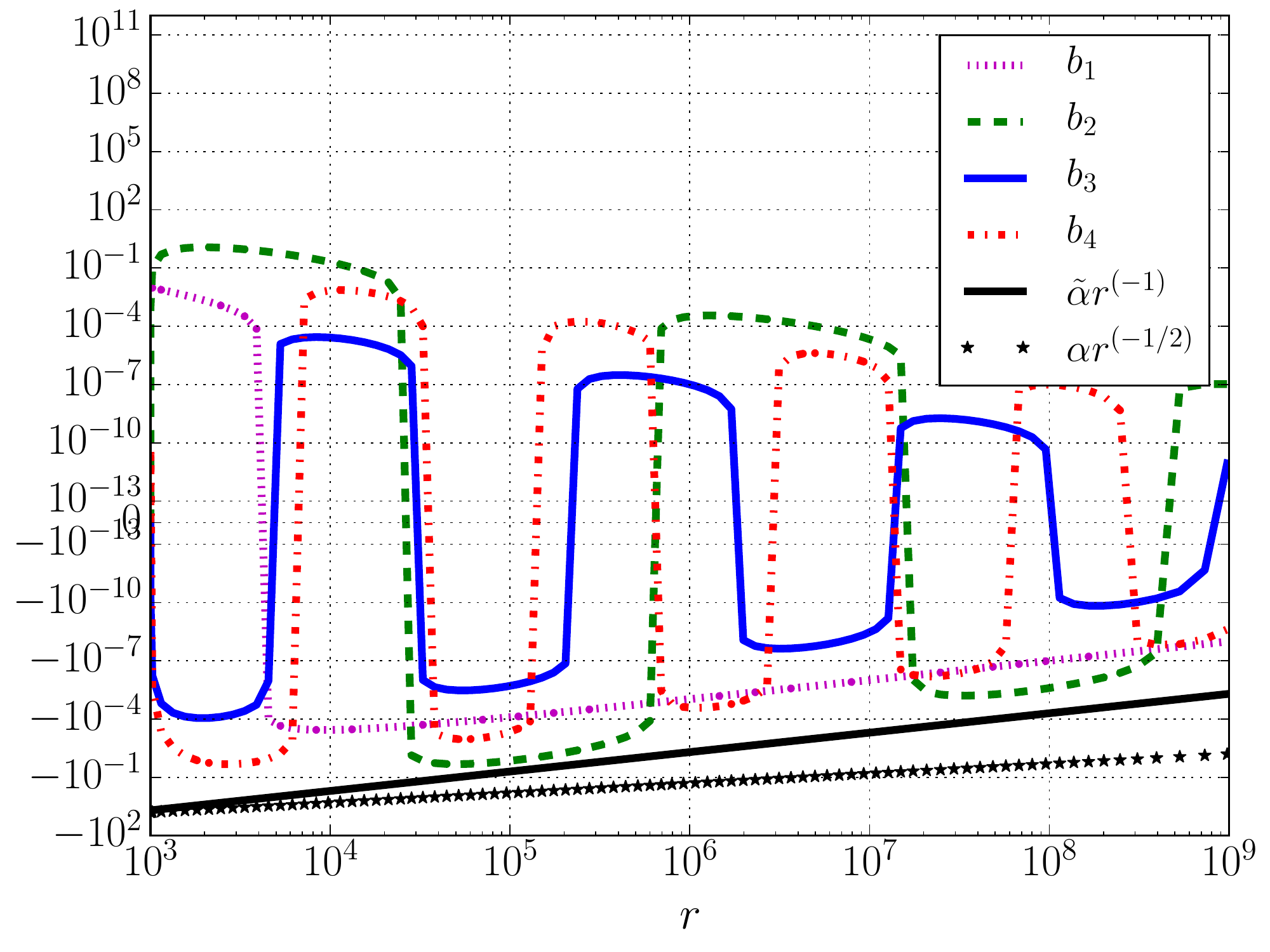}   
     \caption{Numerical evolutions and decay rates for the first modes of $\Bbbkm$ (Case~3).} \label{fig:kmdesNLARGE}
  \end{minipage}      
\end{figure}

\paragraph{Case~3 ($\beta = 10^{-1}$ and $\epsilon = 10^{-2}$).}
Now we choose a relatively large value of $\epsilon$. The heuristic analysis in \Sectionref{sec:heuristics}, for which we assume that $\epsilon$ is small, may therefore not yield an accurate description. Our numerical results are shown in \Figsref{fig:KmodesNLARGE} and \ref{fig:kmdesNLARGE}. It is evident in these plots that $a_0$ is certainly not compatible with the decay rate $r^{-2}$ associated with the strong notion of asymptotic flatness. The dashed curve in \Figref{fig:decaya0} strongly suggests that as Case~1 before,  $a_0$  decays exactly like $r^{-3/2}$ for large values of $r$ and therefore probably also violates the weak notion of asymptotic flatness.

\section{Conclusions}
\label{sec:conclusion}

We have studied the recently proposed hyperbolic/algebraic formulation of the vacuum constraint equations in General Relativity using both heuristic and numerical tools. In particular, 
we studied the asymptotic behavior of solutions of this system which can be viewed as (axisymmetric) perturbations of the Kerr-Schild data for the Schwarzschild space-time. We found that 
these solutions are generically not asymptotically flat in the strong sense of Eqns.~\eqref{eq:AFStrong1}-\eqref{eq:AFStrong2} and we also see indications that they violate the weak 
asymptotic flatness conditions  of Eqns.~(\ref{eq:AFWeak1}-\ref{eq:AFWeak2}).

It is fair to say that up to now no method to construct initial data is known which yields full control of \emph{the actual physics} of the initial data set in a clean and direct manner. People have developed a certain level of intuition and experience for more traditional approaches over many years. But even there, many issues are unsolved (for example, regarding ``junk radiation''). For this relatively new framework considered here, we are only at the beginning of developing this experience.
The goal of our work here was therefore clearly to construct ``simple'', rather than ``physically realistic'' initial data sets. 
%
% , in particular the relatively new framework considered here, it is a consequence of the intriguingly difficult nonlinear coupling between the \emph{actual physics} described by an initial data set and the gauge content that we never really know what the \emph{actual physical content} of the resulting initial data set may be. 
%
% The purpose of the work here is to make the first steps in developing experience for this particular initial data construction procedure.
Moreover, we address only one of many crucial aspects which characterize the physical properties of initial data sets, namely asymptotic flatness. The fact that there is no obvious way to guarantee this property for the approach considered here could be seen as a major drawback.
Building on these first steps, a next natural idea could be to extend our analysis to some other, possible more physically relevant scenario, like Schwarzschild initial data perturbed by a Teukolsky wave.

%Physics ... initial data ... not clear ... gauge ...

%Of course, these findings need to be confirmed and extended. There are several shortcomings in our analysis that should be eliminated. We should also look at the general case without any symmetry,
%even though it seems unlikely that this would change the result dramatically.

One problem  with our present approach is that we start from an arbitrary sphere; there are no geometrically distinguished conditions that one could impose on the data. 
This might be different if one could start the solution process at a single point (a sphere with zero radius) where the conditions of regularity might be enough to fix a unique solution. 
Then the initial data set would be determined essentially by the free data. Another possibility would be to start the solution at infinity and integrate inwards. The conditions of asymptotic 
flatness might be enough to guarantee a unique solution. The big problem in both cases is, however, that one cannot enforce things at the other end: when evolving outwards the solution may 
violate asymptotic flatness and when evolving inwards we may end up with a singular initial data set.

It would be interesting to see whether there are conditions on the free data that could guarantee the correct behavior at the end of the evolution. It would also be interesting to see how things
change when one uses the hyper/parabolic formulation of the constraints instead of the hyperbolic/algebraic formulations that we used in the present article.

%%%%%%%%%%%%%%%%%%%%%%%%%%%%%%%%%%%%%%%%%%%%%%%%%%%%%%%%%%%%%%%%%%%%%%%%%%%%%%%%%%%%%%%%%%%%%%%%%%%%%%%%%%%%%%%%%%%%%%%%%%%%%%%%%%%%%%%%%%%%%%%%%%%%
%-------------------------------------------------------------------------------------------------
\section*{Acknowledgments}
L.E.\ was partly funded by the University of Otago Research Grant ``Dynamical dark energy in the young universe and its consequences 
for the present and future history'' in 2016.

%\bibliographystyle{abbrv}
%\bibliography{bibliography}

\end{document}